\documentclass[12pt]{article}
\usepackage{epsfig}
\usepackage{psfrag}
\usepackage{latexsym}
\usepackage[DIV13]{typearea}
\usepackage{amsmath}
\usepackage{amssymb}
\usepackage{amsfonts}
\usepackage{cite}
\usepackage{bbold}
\usepackage[footnotesize]{caption2}
\usepackage{graphicx}
\usepackage[center,footnotesize,hang]{subfigure}
\usepackage{url}
\usepackage{color}

\def\cL{{\cal L}} \def\cM{{\cal M}}

\def\smu{\sigma^{\mu}}

\def\smub{{\bar\sigma}^{\mu}}

\def\snu{\sigma^{\nu}}

\def\snub{{\bar\sigma}^{\nu}}
\def\smn{\sigma^{\mu\nu}}
\def\smnb{{\bar\sigma}^{\mu\nu}}

\newcommand{\phit}{\varphi_T}

\newcommand{\phis}{\varphi_S}

\newcommand{\Uu}{\text{U(1)}}

\newcommand{\Zt}{\text{Z}_3}

\def\gs{\mathrel{
   \rlap{\raise 0.511ex \hbox{$>$}}{\lower 0.511ex \hbox{$\sim$}}}}
\def\ls{\mathrel{
   \rlap{\raise 0.511ex \hbox{$<$}}{\lower 0.511ex \hbox{$\sim$}}}}

\newcommand{\ba}{\begin{array}{c}}
\newcommand{\baz}{\begin{array}{cc}}
\newcommand{\bad}{\begin{array}{ccc}}
\newcommand{\ea}{\end{array}}

\newcommand{\al}{\alpha}
\newcommand{\be}{\beta}
\newcommand{\ga}{\gamma}
\newcommand{\Ga}{\Gamma}

\newcommand{\La}{\Lambda}
\newcommand{\om}{\omega}

\newcommand{\onep}{1^\prime}
\newcommand{\onepp}{1^{\prime\prime}}

\def\beq{\begin{equation}}
\def\eeq{\end{equation}}
\def\bea{\begin{eqnarray}}
\def\eea{\end{eqnarray}}
\def\bet{\begin{tabular}}
\def\eet{\end{tabular}}
\def\bes{\begin{subequations}\bea}
\def\ees{\eea\end{subequations}}

\def\thh{\theta^2}
\def\thhb{\overline{\theta}^2}
\def\be{\begin{equation}}
\def\ee{\end{equation}}
\def\bc{\begin{center}}
\def\ec{\end{center}}
\def\bea{\begin{eqnarray}}
\def\eea{\end{eqnarray}}
\def\dd{\displaystyle}
\def\nn{\nonumber}

\def\ov{\overline}

\catcode`@=11
\def\marginnote#1{}
\newcount\hour
\newcount\minute
\newtoks\amorpm
\hour=\time\divide\hour by60
\minute=\time{\multiply\hour by60 \global\advance\minute by-\hour}
\edef\standardtime{{\ifnum\hour<12 \global\amorpm={am}%
        \else\global\amorpm={pm}\advance\hour by-12 \fi
        \ifnum\hour=0 \hour=12 \fi
        \number\hour:\ifnum\minute<10 0\fi\number\minute\the\amorpm}}
\edef\militarytime{\number\hour:\ifnum\minute<10 0\fi\number\minute}
\def\draftlabel#1{{\@bsphack\if@filesw {\let\thepage\relax
   \xdef\@gtempa{\write\@auxout{\string
      \newlabel{#1}{{\@currentlabel}{\thepage}}}}}\@gtempa
   \if@nobreak \ifvmode\nobreak\fi\fi\fi\@esphack}
        \gdef\@eqnlabel{#1}}
\def\@eqnlabel{}
\def\@vacuum{}
\def\draftmarginnote#1{\marginpar{\raggedright\scriptsize\tt#1}}
\def\draft{\oddsidemargin 0.0truein
        \def\@oddfoot{\sl preliminary draft \hfil
        \rm\thepage\hfil\sl\today\quad\militarytime}
        \let\@evenfoot\@oddfoot \overfullrule 3pt
        \let\label=\draftlabel
        \let\marginnote=\draftmarginnote
   \def\@eqnnum{(\theequation)\rlap{\kern\marginparsep\tt\@eqnlabel}%
\global\let\@eqnlabel\@vacuum}  }
\catcode`@=12

\begin{document}
\begin{titlepage}
\vspace*{-1cm}
\phantom{hep-ph/***}
\hfil{SISSA 37/2009/EP}\hfill{DFPD-09/TH/13}\\
\vskip 2.5cm
\begin{center}
{\Large\bf Lepton Flavour Violation in a Supersymmetric Model \\
\vskip .3cm
with $A_4$ Flavour Symmetry}
\end{center}
\vskip 0.5  cm
\begin{center}
{\large Ferruccio Feruglio}~$^{a)}$\footnote{e-mail address: feruglio@pd.infn.it},
{\large Claudia Hagedorn}~$^{b)}$\footnote{e-mail address: hagedorn@sissa.it},
\\
\vskip .2cm
{\large Yin Lin}~$^{a)}$\footnote{e-mail address: yin.lin@pd.infn.it} and
{\large Luca Merlo}~$^{a)}$\footnote{e-mail address: merlo@pd.infn.it}
\\
\vskip .2cm
$^{a)}$~Dipartimento di Fisica `G.~Galilei', Universit\`a di Padova
\\
INFN, Sezione di Padova, Via Marzolo~8, I-35131 Padua, Italy
\\
\vskip .1cm
$^{b)}$~
SISSA, Scuola Internazionale Superiore di Studi Avanzati\\
Via Beirut 2-4, I-34014 Trieste, Italy\\
and\\ INFN, Sezione di Trieste, Italy
\end{center}
\vskip 0.2cm
\begin{abstract}
\noindent
We compute the branching ratios for $\mu\to e \gamma$, $\tau\to\mu\gamma$ and $\tau\to e \gamma$ in a supersymmetric model
invariant under the flavour symmetry group $A_4\times Z_3\times U(1)_{FN}$, in which near tri-bimaximal lepton mixing
is naturally predicted. 
At leading order in the small symmetry breaking parameter $u$, which is of the same order as the reactor mixing angle
$\theta_{13}$, we find that the branching ratios generically scale as $u^2$. 
Applying the current bound on the branching ratio of $\mu\to e\gamma$ shows that small values of $u$ or $\tan\beta$ 
are preferred in the model for mass parameters $m_{SUSY}$ and $m_{1/2}$ smaller than $1000$ GeV. 
The bound expected from the on-going MEG experiment will provide a severe constraint on the 
parameter space of the model either enforcing $u \approx 0.01$ and small $\tan\beta$ or $m_{SUSY}$ and $m_{1/2}$ above
$1000$ GeV. In the special case of universal soft supersymmetry breaking terms
in the flavon sector a cancellation takes place in the amplitudes and the branching ratios scale as $u^4$, allowing for smaller slepton masses.
The branching ratios for $\tau \to \mu \gamma$ and $\tau\to e \gamma$ are predicted to be of the same order as the one for $\mu\to e \gamma$, 
which precludes the possibility of observing these $\tau$ decays in the near future. 
\end{abstract}
\end{titlepage}
\setcounter{footnote}{0}
\vskip2truecm


\section{Introduction}
Flavour violation in the lepton sector (LFV) has been firmly established in neutrino oscillations.
A direct evidence of conversion of electron neutrinos into a combination of muon and tau neutrinos
is provided by solar neutrino oscillations. In a three neutrino framework atmospheric neutrino data
can only be explained by the dominant oscillation of muon neutrinos into tau neutrinos.
It is natural to expect that LFV takes place, at least at some level, in other processes such as
those involving charged leptons. Flavour violating decays of charged leptons, strictly forbidden
in the Standard Model (SM), are indeed allowed as soon as neutrino mass terms are considered. If neutrino masses are the only source of LFV,
the effects are too small to be detected, but in most extensions of the SM in which new particles and new interactions
with a characteristic scale $M$ are included, the presence of new sources of flavour violation,
in particular in the lepton sector, is a generic feature.  In a low-energy description, the corresponding effects can be parametrized
by higher-dimensional operators describing flavour-violating rare decays of the charged leptons.
The dominant terms are represented by dimension six operators, suppressed by two powers of $M$.
The present bounds on the branching ratios \cite{Raidal} of these decays set stringent limits on combinations
of the scale $M$ and the coefficients of the involved operators. Typically, for coefficients of order one, the existing bounds
require a large scale $M$, several orders of magnitude larger than the TeV scale.
Conversely, to allow for new physics at the TeV scale, coefficients much smaller than one
are required, which might suggest the presence of a flavour symmetry.

New physics at the TeV scale supplemented by a flavour symmetry provides an interesting framework.
Flavour symmetries have been invoked to describe the observed pattern of lepton masses and mixing angles \cite{Data},
but quite often this approach is limited to a fit of the existing data, with very few new
testable predictions \cite{reviewaf}. Specific relations among LFV processes are usually consequences
of flavour symmetries and of their pattern of symmetry breaking. This allows to get
independent information on the flavour symmetry in the charged lepton sector. Moreover, if $M$ is sufficiently small,
new particles might be produced and detected at the LHC, with features that could additionally
confirm or reject the assumed symmetry pattern.
In \cite{A4eff} we have recently analyzed in an effective Lagrangian approach LFV processes within a class of models
possessing  a flavour symmetry $G_f=A_4\times Z_3\times U(1)_{FN}$. Models of this class \cite{af1,af2,afl,afh}
automatically reproduce nearly tri-bimaximal (TB) lepton mixing at the leading order (LO) via a vacuum alignment mechanism.
\footnote{Due to the success of these models in the lepton sector several extensions to the quark sector can be found in the literature \cite{QuarkExtention}.}
In this type of models \cite{TBother,TBA4,af1,af2,afl,afh}, corrections to TB mixing are generically proportional to the symmetry breaking parameter $u$,
especially the reactor mixing angle $\theta_{13}$ is of the order of $u$.\footnote{For alternative scenarios see \cite{AFM_S4,Lin_LargeReactor}.} 
The parameter $u$ is expected to vary between a few per mil and a few percent.
In particular we have evaluated the normalized branching ratios $R_{ij}$ for the LFV transitions
$l_i\to l_j \gamma$:
\be
R_{ij}=\frac{BR(l_i\to l_j\gamma)}{BR(l_i\to l_j\nu_i{\bar \nu_j})}~~~.
\ee
In \cite{A4eff} we found that the generic expectation for the ratios $R_{ij}$ is:
\be
R_{ij}=\frac{48\pi^3 \alpha}{G_F^2 M^4}\vert w_{ij} ~u\vert^2
\label{LFV}
\ee
where $\alpha$ is the fine structure constant, $G_F$ is the Fermi constant and
$w_{ij}$ are dimensionless parameters that cannot be predicted within the effective Lagrangian approach, but are
expected to be of order one. Given the range of  $u$, it turns out that
the present bound on $\mu\to e \gamma$ requires the scale $M$ to be larger than about 10 TeV.
If the underlying fundamental theory is weakly interacting, this translates into a lower bound of about $g M/(4\pi)$
on the typical mass of the new particles. 
At variance with this generic estimate, it was also observed that, if the flavour symmetry is realized in a supersymmetric (SUSY) model,
a cancellation might take place in the amplitudes for LFV transitions. An argument was given suggesting that in this case the ratios $R^{SUSY}_{ij}$
are of the form:
\be
R^{SUSY}_{ij}= \frac{48\pi^3 \alpha}{G_F^2 M^4}\left[\vert w^{(1)}_{ij} u^2\vert^2+\frac{m_j^2}{m_i^2} \vert w^{(2)}_{ij} u\vert^2\right]
\label{LFVsusy}
\ee
where $M$ now corresponds to the supersymmetry breaking scale, $m_i$ ($i=e,\mu,\tau$) are the charged lepton masses
and $w^{(1,2)}_{ij}$ are unknown dimensionless quantities of order one. Therefore in a supersymmetric context
the predicted rates are more suppressed, allowing for new physics close to the TeV
scale, without conflicting with the present bounds. Since the cancellation expected in the supersymmetric case
considerably changes the conclusion that can be derived from the existing bound on $\mu\to e \gamma$,
it is important to perform a direct computation of the branching ratios within an explicit supersymmetric model
incorporating the flavour symmetry $A_4\times Z_3\times U(1)_{FN}$.

In the present paper we consider the $A_4$ realization proposed in \cite{A4eff} in which SUSY breaking effects were ignored
and we extend it to a more realistic model, by adding a full set of SUSY breaking terms consistent with the
flavour symmetry. We assume that the breaking of supersymmetry occurs at a scale higher than or comparable to the flavour scale,
simulated in our effective Lagrangian by a cutoff, so that at energies close to the cutoff scale we have non-universal
boundary conditions for the soft SUSY breaking terms, dictated by the flavour symmetry. Depending on the assumed mechanism of SUSY breaking
we may have boundary conditions different from these, possibly enforced at a smaller energy scale. For this reason,
our approach maximizes the possible effects on LFV processes.

Through a detailed calculation of the slepton mass matrices in the physical basis,
in which kinetic terms are canonical and leptons are in their mass eigenbasis, and evaluation of $R_{ij}$ in the mass insertion (MI) approximation
we find that the behaviour of $R_{ij}$, expected from the SUSY variant of the effective Lagrangian approach, 
given in eq. (\ref{LFVsusy}) is violated by a single, flavour independent
contribution. The correct scaling of $R_{ij}$ is the one of eq. (\ref{LFV}), with a universal constant $w_{ij}$.
This implies $R_{\mu e}=R_{\tau\mu}=R_{\tau e}$ at the LO in $u$.
We identify the source of violation of the expected behaviour in a contribution to the right-left (RL) block of the slepton mass matrix,
associated to the sector necessary to maintain the correct breaking of the flavour symmetry $A_4$. We also enumerate the conditions
under which such a contribution is absent and the behavior in eq. (\ref{LFVsusy}) is recovered, though we could not find
a dynamical explanation to justify the realization of these conditions in our model.

As shown in the MI approximation the coefficient $w_{ij}$ of the leading contribution 
in eq. (\ref{LFV}) is numerically small in a large region of the parameter space,
where the sub-leading contributions play an important role. 
We then provide a numerical study of $R_{ij}$ by using full one-loop expressions and explore the parameter space of the model.
Thereby, we assume a supergravity (SUGRA) framework with a common mass scale $m_{SUSY}$ for soft sfermion and Higgs masses 
and a common mass $m_{1/2}$ for gauginos at high energies.
Applying the current MEGA bound on $R_{\mu e}$ we find that small values of $u$ or 
$\tan\beta$ are favoured for $m_{SUSY}$ and $m_{1/2}$ below $1000$ GeV. Employing the foreseen bound coming from the
MEG experiment constrains the model severely allowing only for small $u$ and $\tan\beta$ for $m_{SUSY}, m_{1/2} \lesssim 1000$ GeV
or requires larger $m_{SUSY}$ and $m_{1/2}$ reducing the prospects for detection of sparticles at LHC.
 Furthermore, it turns out to be rather unnatural to reconcile the values of superparticle masses necessary to
account for the measured deviation $\delta a_\mu$ in $a_\mu$, the muon anomalous magnetic moment, from the SM value
with the present bound on $R_{\mu e}$.
In our model values of $\delta a_{\mu}$ smaller than $100 \times 10^{-11}$ are favoured.

The paper is structured as follows: in section 2 we review the basic features of the SUSY $A_4$ model and calculate kinetic
terms and mass matrices of leptons and sleptons in an expansion in the symmetry breaking parameters. In section 3 we compute
the slepton mass matrices in the physical basis.
Furthermore, we estimate the possible effects on the slepton masses coming from renormalization group (RG) running. Section 4 is dedicated
to the study of the quantities $R_{ij}$ in the MI approximation, while in section 5 we discuss $R_{\mu e}$ in a more
quantitative way in the SUGRA context and comment on the size of the measured deviation in the muon anomalous magnetic moment $a_\mu$
in our model. Finally, we conclude in section 6. Appendix A contains
details about the group theory of $A_4$, appendix B details of the calculation of the slepton mass matrices in the physical basis
and in appendix C conventions and formulae, used in the numerical study of section 5, are given.


\section{A SUSY model with $A_4$ flavour symmetry}

In this section, we first discuss the main features and predictions of the lepton mixing
in the class of $A_4$ realizations to which the present SUSY model belongs.
Then we will present the K\"{a}hler and the superpotential of our model
from which kinetic and mass terms are derived for leptons and sleptons.


\subsection{General features}

The flavour symmetry $A_4$ can give rise to TB mixing in the lepton sector, if it is broken in a specific way, as discussed in
\cite{TBA4,af1,af2,afl,afh}. The group $A_4$ is the group of even permutations of four objects which is isomorphic to the tetrahedral group.
It has 12 elements and four irreducible representations: three inequivalent singlets, denoted by $1$, $1'$ and $1''$ and a triplet $3$.
$A_4$ is generated by two generators $S$ and $T$ which fulfill the following relations \cite{A4Presentation}:
\be
S^2=(ST)^3=T^3=1~~~.
\label{$A_4$}
\ee
A set of generators $S$ and $T$ for all representations as well as Kronecker products and Clebsch Gordan coefficients are given
in appendix A. The specific breaking pattern of $A_4$ which leads to TB mixing for leptons requires that a $Z_3$ subgroup, called $G_T$,
is preserved in the charged lepton sector, whereas a $Z_2$ group, $G_S$, remains unbroken in the neutrino sector.
The $Z_3$ group is generated by the element $T$ and is left intact, if a scalar triplet $\varphi=(\varphi_1,\varphi_2,\varphi_3)$
acquires a vacuum expectation value (VEV) of the form:
\be
\langle\varphi\rangle \propto (1,0,0) \; .
\label{unozero}
\ee
Breaking $A_4$ down to $G_S$ is achieved by a triplet which gets a VEV:
\be
\langle\varphi\rangle \propto (1,1,1) \; .
\label{unotre}
\ee
Apart from $A_4$, being responsible for TB mixing, the theory is invariant under additional flavour symmetries
so that the full symmetry is:
\be
G_f=A_4\times Z_3\times \Uu_{FN}\times \Uu_{R} \; .
\label{flavgroup}
\ee
The cyclic symmetry $Z_3$ (not to be confused with the $Z_3$ subgroup of $A_4$ preserved at LO in the charged lepton sector)
is necessary in order to avoid large mixing effects between the flavons that give masses to the charged leptons and those giving masses to neutrinos.
The mass hierarchy among the charged leptons is explained through the $\Uu_{FN}$ factor \cite{fn}. The second $\Uu$ factor, $\Uu_{R}$,
is a continuous $R$-symmetry containing $R$-parity as a subgroup. The flavour symmetry $G_f$ is broken by
the chiral superfields $\varphi_T$, $\varphi_S$, $\xi$ and $\theta_{FN}$ whose transformation properties under $G_f$
are shown in table 1 together with those of the lepton
supermultiplets $l$, $e^c$, $\mu^c$, $\tau^c$ and of the electroweak doublets $H_{u,d}$.
Introducing the driving fields $\varphi^T_0$, $\varphi^S_0$ and $\xi_0$, see also table 1, we can write down the following 
superpotential: \footnote{The additional flavon field $\tilde\xi$ transforms in the same way under the symmetries
of the model as the field $\xi$. However, it does not acquire a VEV at LO and thus is not relevant for our discussion here.
For further details see \cite{af2}.}
\begin{eqnarray}
w_d&=&M_T (\varphi_0^T \varphi_T)+ g (\varphi_0^T \varphi_T\varphi_T)\nn\\
&+&g_1 (\varphi_0^S \varphi_S\varphi_S)+
g_2 \tilde{\xi} (\varphi_0^S \varphi_S)+
g_3 \xi_0 (\varphi_S\varphi_S)+
g_4 \xi_0 \xi^2+
g_5 \xi_0 \xi \tilde{\xi}+
g_6 \xi_0 \tilde{\xi}^2~~~ +....
\label{wd}
\end{eqnarray}
where dots denote sub-leading non-renormalizable corrections.
By $(\cdots)$ we denote the contraction to an $A_4$ invariant.
In the limit of unbroken supersymmetry all $F$ terms vanish. From these conditions we can derive the vacuum
alignment of the flavons $\varphi_T$, $\varphi_S$ and $\xi$: 
\bea
\frac{\langle\varphi_T\rangle}{\Lambda_f}&=&(u,0,0)+(c' u^2,c u^2,c u^2)+O(u^3)\nonumber\\
\label{vevs}
\frac{\langle\varphi_S\rangle}{\Lambda_f}&=& c_b(u,u,u)+O(u^2)\\
\frac{\langle\xi\rangle}{\Lambda_f}&=&c_a u+O(u^2)\nonumber
\eea
where $c, c',c_{a,b}$ are complex numbers with absolute value of order one and 
$u$ is one of the two small symmetry breaking parameters in the theory.
The scale $\La_f$ is the cutoff scale associated to the flavour symmetry. Its value is expected to be around the scale of grand
unification $10^{16}$ GeV.
We display the sub-leading $O(u^2)$ corrections, which arise from non-renormalizable terms in the superpotential
$w_d$ and which have been studied in detail in \cite{af2}, only for the field $\varphi_T$, since only these are relevant in the
following analysis, whereas those of the fields $\xi$ and $\varphi_S$ are important in the computation of neutrino masses and lepton mixings.
We remark that in the SUSY limit the VEVs of the driving fields $\varphi^T_0$, $\varphi^S_0$ and $\xi_0$ are zero. This
is however in general no longer true, if we include soft SUSY breaking terms into the flavon potential, as we shall see in the next section
and as has been discussed in \cite{VEVsaux}.

For the Froggatt-Nielsen (FN) field $\theta_{FN}$ to acquire a VEV, we assume that the symmetry $\Uu_{FN}$ is gauged such that 
$\theta_{FN}$ gets its VEV through a $D$ term. The corresponding potential is of the form:
\begin{equation}
\label{Dterm}
V_{D, FN}=\frac{1}{2}(M_{FI}^2- g_{FN}\vert\theta_{FN}\vert^2+...)^2
\end{equation}
where $g_{FN}$ is the gauge coupling constant of $\Uu_{FN}$ and $M_{FI}^2$ denotes the contribution of the Fayet-Iliopoulos (FI) term. 
Dots in eq.(\ref{Dterm}) represent e.g. terms involving the right-handed charged leptons $e^c$ and $\mu^c$ 
which are charged under $\Uu_{FN}$. These terms
are relevant in the calculation of the different contributions to the soft mass terms of right-right (RR) type, as we shall see below.
$V_{D,FN}$ leads in the SUSY limit to:
\begin{equation}
|\langle\theta_{FN}\rangle|^2= \frac{M_{FI}^2}{g_{FN}}
\end{equation}
which we parametrize as:
\begin{equation}
\label{vevtheta}
\frac{\langle\theta_{FN}\rangle}{\Lambda_f}=t
\end{equation}
with $t$ being the second small symmetry breaking parameter in our model. Both
$u$ and $t$ can also be in general complex, but through field redefinitions of the supermultiplets $\varphi_T$ and $\theta_{FN}$, they can
be made real and positive.
\begin{table}[!ht]
\centering
                \begin{math}
                \begin{array}{|c||c|c|c|c||c|c|c|c|c||c|c|c|}
                    \hline
                    &&&&&&&&&&&& \\[-9pt]
                    \text{Field} & l & e^c & \mu^c & \tau^c & H_{u,d} & \phit & \phis & \xi & \theta_{FN} & \varphi^T_0 & \varphi^S_0 & \xi_0\\[10pt]
                    \hline
                    &&&&&&&&&&&&\\[-9pt]
                    A_4 & 3 & 1 & \onepp & \onep & 1 & 3 & 3 & 1 &  1 & 3 & 3 & 1\\[3pt]
                    \hline
                    &&&&&&&&&&&&\\[-9pt]
                    \Zt & \om & \om^2 & \om^2 & \om^2  & 1 & 1& \om & \om  & 1 & 1 & \om & \om\\[3pt]
                    \hline
                    &&&&&&&&&&&&\\[-9pt]
                    \Uu_{FN} & 0 & 2 & 1 & 0  & 0 & 0 & 0 & 0 & -1 & 0 & 0 & 0\\[3pt]
                    \hline
                    &&&&&&&&&&&&\\[-9pt]
                    \Uu_{R} & 1 & 1 & 1 & 1  & 0 & 0 & 0 & 0 & 0 & 2 & 2 & 2\\[3pt]
                    \hline
                \end{array}
               \end{math}
            \caption{ Transformation properties of the fields
            under \rm{$A_4$}, \rm{$\Zt$}, \rm{$\Uu_{FN}$} and \rm{$\Uu_{R}$}.}
            \end{table}
\vskip 0.2cm
\noindent
\vskip 0.2cm
\noindent At the LO, the mass matrices of charged leptons and light neutrinos have the following form:
\be
m_l \propto \left(
\begin{array}{ccc}
y_e t^2 & 0& 0\\
0& y_\mu t& 0\\
0& 0& y_\tau
\end{array}
\right) u~~~,
\label{yf}
\ee
and
\be
m_\nu \propto \left(
\begin{array}{ccc}
a+2 b/3& -b/3& -b/3\\
-b/3& 2b/3& a-b/3\\
-b/3& a-b/3& 2 b/3
\end{array}
\right) u~~~,
\label{Y}
\ee
with $y_e$, $y_\mu$, $y_\tau$, $a$ and $b$ being complex numbers with absolute values of order one.
As one can see the relative hierarchy among the charged lepton masses is given by the parameter $t$ for which we
take:
\be
t\approx 0.05~~~.
\label{tbound}
\ee
The neutrino mass matrix is diagonalized by:
\be
U_{TB}^T m_\nu U_{TB}  \propto {\tt diag}(a+b,a,-a+b) u~~~,
\ee
where $U_{TB}$, up to phases, is the TB mixing matrix\cite{TB}:
\be
U_{TB}=\left(
\begin{array}{ccc}
\sqrt{2/3}& 1/\sqrt{3}& 0\\
-1/\sqrt{6}& 1/\sqrt{3}& -1/\sqrt{2}\\
-1/\sqrt{6}& 1/\sqrt{3}& +1/\sqrt{2}
\end{array}
\right)~~~.
\label{UTB}
\ee
The allowed range of the parameter $u$ is determined by the requirement that sub-leading corrections which perturb the
LO result are not too large and by the requirement that the $\tau$ Yukawa coupling $y_\tau$ does not become too large.
The first requirement results in an upper bound on $u$ of about $0.05$, which mainly comes from the fact that the solar
mixing angle should remain in its $1\sigma$ range \cite{Data}. The second one gives a lower bound which we estimate as:
\be
u=\dd\frac{\tan\beta}{|y_\tau|} \dd\frac{\sqrt{2} m_\tau}{v} \approx 0.01 \dd\frac{\tan\beta}{|y_\tau|}
\label{tanb&u&yt}
\ee
where $v\approx 246$ GeV and $\tan\beta$ is the ratio between the VEVs of the neutral spin zero components
of $H_u$ and $H_d$, the two doublets responsible for electroweak symmetry breaking. For the $\tau$ lepton we use its pole mass $m_\tau=(1776.84 \pm 0.17) \;
\rm{MeV}$ \cite{pdg2008}. Requesting $|y_\tau|<3$ we find a lower limit on $u$
close to the upper bound $0.05$ for $\tan\beta=15$, whereas $\tan\beta=2$ 
\footnote{It is known that $\tan\beta$ cannot be too small \cite{tanbetabound}. Here we take 
$\tan\beta=2$ as the smallest allowed value.}
gives as lower limit $u \gtrsim 0.007$.
Obviously, these limits depend on the largest allowed value of $|y_\tau|$, as well as on whether we
identify $m_\tau$ with the pole mass or with the $\tau$ mass renormalized at some reference scale, such as the scale of grand unification.
We choose as maximal range:
\be
0.007 \lesssim u \lesssim 0.05~~~,
\label{ubound}
\ee
which shrinks when $\tan\beta$ is increased from 2 to 15. 
Concerning the relative size of the two symmetry breaking parameters we note that $u \lesssim t$ holds for all values of $u$.
\vskip 0.5 cm


\subsection{The SUSY Lagrangian and the soft SUSY breaking terms}
\vskip 0.5 cm
We analyse the Lagrangian of the model:
\bea
{\cal L}&=&\int d^2\theta d^2\overline{\theta} {\cal K}(\ov{z}, e^{2 V} z)+\left[\int d^2 \theta w(z)+h.c.\right]\nn\\
&+&\frac{1}{4}\left[\int d^2\theta f(z) {\cal W W}+h.c.\right]~~~,
\label{leel}
\eea
where ${\cal K}(\ov{z},z)$ is the K\"ahler potential, a real gauge-invariant function of the chiral superfields $z$ and their conjugates, of dimensionality
(mass)$^2$; $w(z)$ is the superpotential, an analytic gauge-invariant function of the chiral superfields, of dimensionality (mass)$^3$;
$f(z)$ is the gauge kinetic function, a dimensionless analytic gauge-invariant function; $V$ is the Lie-algebra valued vector supermultiplet,
describing the gauge fields and their superpartners. Finally ${\cal W}$ is the chiral superfield describing, together with the function $f(z)$,
the kinetic terms of gauge bosons and their superpartners.
Each of the terms on the right-hand side can be written in an expansion in powers of the flavon fields. Since we have two independent
symmetry breaking parameters, $u$ and $t$, see eqs. (\ref{vevs}) and eq. (\ref{vevtheta}), we consider a double expansion of ${\cal L}$ in powers of $u$ and $t$.
In this expansion we keep terms up to the second order in $u$, i.e. terms quadratic in the fields $\varphi_{S,T}$ and $\xi$.
The expansion in the parameter $t$, responsible for the breaking of the Froggatt-Nielsen U(1)$_{FN}$ symmetry, is stopped
at the first non-trivial order, that is by allowing as many powers of the field $\theta_{FN}$ as necessary in order to
obtain non-vanishing values for all entries of the matrices describing lepton masses as well as for the entries of
the matrices describing kinetic terms and slepton masses.
\footnote{Concerning the K\"ahler potential we observe that we can additionally write down operators involving the total invariant
$\overline{\theta}_{FN} \theta_{FN}=|\theta_{FN}|^{2}$. These contribute to the diagonal elements of
the kinetic terms and the slepton masses. In the K\"ahler potential for the left-handed fields they can be safely neglected, since the
LO correction is of $O(u)$. In the right-handed sector, they contribute at the
same order as the terms arising through a double flavon insertion.}
Finally, second order corrections in $u$ also arise from the sub-leading terms of the VEV $\langle\varphi_T\rangle$, eq. (\ref{vevs}),
and are included in our estimates. 

The soft SUSY breaking terms are generated from the SUSY Lagrangian by promoting all coupling constants, such as Yukawa couplings,
couplings in the flavon superpotential and couplings in the K\"ahler potential, to superfields with constant $\thh$ and $\thh\thhb$ components 
\cite{LutyReview}.
Through this we derive subsequently the soft masses $(m_{(e,\nu) LL}^2)_K$ and $(m_{e RR}^2)_K$ from the K\"ahler potential.
One contribution to $m_{eRL}^2$, which we call $(m_{e RL}^2)_1$ in the following, arises from the Yukawa couplings present in the superpotential
$w$. 

Important contributions to slepton masses originate from the modification of the VEVs of flavons and driving fields 
due to SUSY breaking effects. 
A detailed study of the VEVs of these fields and their dependence on the soft SUSY breaking parameters
is presented in \cite{VEVsaux} and we summarize the main results here.
When soft SUSY breaking terms are included into the flavon potential, the VEVs in eq. (\ref{vevs}) receive
additional contributions of order $m_{SUSY}$, completely negligible compared to $\Lambda_f~u$. At the same time,
the driving fields $\varphi^T_0$, $\varphi^S_0$ and $\xi_0$ develop a VEV of the size of the soft SUSY breaking scale $m_{SUSY}$. 
An equivalent statement is that the auxiliary fields of the flavons acquire a VEV at the LO of the size of $m_{SUSY} \times u \, \La_f$.
Especially, for the auxiliary fields contained in the flavon supermultiplet $\varphi_T$ we have \cite{VEVsaux}:
\begin{equation}
\frac{1}{\La_f} \left\langle \frac{\partial w}{\partial \varphi_T} \right\rangle = \zeta \, m_{SUSY} \, \left\{ (u,0,0) 
+ (c_F^\prime u^2, c_F u^2, c_F u^2) \right\} 
\label{VEVsauxphiT}
\end{equation}
where $\zeta$, $c_F^\prime$ and $c_F$ are in general complex numbers with absolute value of order one. 
The parameter $\zeta$ vanishes in the special case of universal soft mass terms in the flavon potential.
When different from zero, the VEVs of the auxiliary components of the flavon supermultiplet
$\varphi_T$ generate another contribution to the soft masses of RL-type, which we denote as $(m_{e RL}^2)_2$.
This contribution is analogous to the one which has been found before
in the supergravity context and which can have a considerable effect on the size of the branching ratio of radiative leptonic
decays, as shown in \cite{FtermSUGRA1,FtermSUGRA2}. Indeed, as we shall see below, in the global SUSY model under consideration
the leading dependence of the normalized branching ratios $R_{ij}$ on $u$ is dominated by $(m_{e RL}^2)_2$.
We remark that the VEVs in eq. (\ref{VEVsauxphiT}) and those of the corresponding flavon field $\varphi_T$ in eq. (\ref{vevs}) have a similar structure but they are not 
proportional, in general. This is due to the different coefficients $c$, $c^\prime$ and $c_F$, $c_F^\prime$, which can be qualitatively understood as follows:
the coefficients $c$, $c^\prime$ mainly depend on a set of parameters that remain in the SUSY limit and receive completely negligible corrections 
from the SUSY breaking terms. On the contrary $\langle \partial w/\partial \varphi_T\rangle$ vanishes in the SUSY limit, to all orders in $u$, 
and $c_F$, $c_F^\prime$ crucially depend on the set of parameters describing the SUSY breaking. We will see that,
if $c$ and $c_F$ accidentally coincide (up to complex conjugation), a cancellation in the leading behaviour of $R_{ij}$ takes place. 

Similarly, the VEV of the FN field $\theta_{FN}$ becomes
shifted, when soft SUSY breaking terms are included into the potential, so that: 
\begin{equation}
\frac{M_{FI}^2}{g_{FN}} - |\langle\theta_{FN}\rangle|^2 = c_{\theta} \, m_{SUSY}^2 \; ,
\end{equation}
with $c_{\theta}$ being an order one number, holds. 
This will lead to a contribution $(m_{e RR}^2)_{D,FN}$  to the soft masses of RR-type, 
since only the right-handed charged leptons $e^c$ and $\mu^c$ are 
charged under $\Uu_{FN}$. Apart from these there are supersymmetric contributions to $m_{(e,\nu) LL}^2$ and $m_{e RR}^2$ from $F$ and $D$ terms,
$(m_{(e,\nu) LL}^2)_{F (D)}$ and $(m_{e RR}^2)_{F (D)}$,
as well as a contribution to $m_{e RL}^2$ coming from the $F$ term of $H_d$, called $(m_{e RL}^2)_{3}$ in the following.

In our notation a chiral superfield and its $R$-parity even
component are denoted by the same letter. The $R$-parity odd component is indicated by a tilde in the following and the conjugate 
(anti-chiral) superfield is denoted by a bar.


\subsubsection{K\"ahler potential}
The expansion of the K\"ahler potential can be written as:
\be
{\cal K}={\cal K}^{(0)}+{\cal K}^{(1)}+{\cal K}^{(2)}+...
\ee
In our model the K\"ahler potential deviates from the canonical form, ${\cal K}(\bar{z},z)=\bar{z}  z$, due to the contributions of non-renormalizable terms,
invariant under both the gauge and the flavour symmetries. Such contributions are sub-leading in the $(u,t)$ expansion, but in principle they are non-negligible,
since the redefinitions required to arrive at canonically normalized fields can also affect lepton and slepton mass matrices.
The LO term in $u$ is given by:
\be
{\cal K}^{(0)}=k_0\sum_{i=1}^3\bar{l}_i l_i+ \sum_{i=1}^3 \left[(k_0^c)_i +({\hat k}^c_0)_i\dd\frac{\vert\theta_{FN}\vert^2}{\Lambda_f^2}\right]\bar{l^c}_i  l^c_i+k_u \vert H_u\vert^2+k_d \vert H_d\vert^2+ k_{FN} \vert  \theta_{FN} \vert^2 + ...
\label{kappa0}
\ee
where $l^c=(e^c,\mu^c,\tau^c)$ and dots stand for additional contributions related to the flavon sector.
The quantities $k_0$, $(k_0^c)_i$, $({\hat k}^c_0)_i$, $k_{u,d}$, $k_{FN}$ and the corresponding ones in the following formulae, are treated as superfields with a constant $\thh\thhb$ component,
as is explained above. Concerning the effect of the superfield $\theta_{FN}$ on the K\"ahler potential ${\cal K}^{(0)}$
of the supermultiplets $l_i$ and $l^c_i$
note that for our purposes we can neglect such terms, with the exception of
the right-handed sector, where terms up to the second order in $t$ have to be taken into account.

At the first order in $u$ we have:
\be
{\cal K}^{(1)}=\dd\frac{k_S}{\Lambda_f}  (\varphi_T (\bar{l}  l)_S)+\dd\frac{k_A}{\Lambda_f}  (\varphi_T (\bar{l}  l)_A)+\dd\frac{k'_S}{\Lambda_f}  (\ov{\varphi}_T (\bar{l}  l)_S)+\dd\frac{k'_A}{\Lambda_f}  (\ov{\varphi}_T (\bar{l}  l)_A)+h.c.
\ee
$(\cdot\cdot\cdot)$ denotes an invariant under $A_4$, while $(\cdot\cdot\cdot)'$ and $(\cdot\cdot\cdot)''$ stand for the $1'$ and $1''$ singlets. Finally two $A_4$ triplets, $a$ and $b$, can be combined into the symmetric triplet $(a b)_S$ or the anti-symmetric one $(a b)_A$. The SU(2) singlet fields $l^c$ are not affected by the first-order correction ${\cal K}^{(1)}$.

At the second order in $u$ we have a richer structure:
\be
{\cal K}^{(2)}={\cal K}_L^{(2)}+{\cal K}_R^{(2)}~~~,
\ee
with the labels $L$ and $R$ referring to lepton doublets $l$ and singlets $l^c$, respectively.
For lepton doublets we find:
\be
{\cal K}_L^{(2)}= \sum_{i=1}^7 \dd\frac{k_i}{\Lambda_f^2}  (X_i \bar{l}  l)
\label{here}
\ee
where $X$ is the list of $Z_3$-invariant operators, bilinear in the flavon superfields $\varphi_{S,T}$ and $\xi$ and their conjugates,
\be
X=\left\{\ov{\xi} \xi,\varphi_T^2,(\ov{\varphi}_T)^2,{\ov \varphi_T}\varphi_T,\ov{\varphi}_S \varphi_S,\ov{\xi}\varphi_S,\ov{\varphi}_S\xi\right\}~~~,
\label{Xlist}
\ee
and each quantity $k_i$ represents a list of parameters since there can be different non-equivalent ways of combining $X_i$ with $\ov{l} l$ to form
an $A_4$-invariant.
There are also obvious relations among the coefficients $k_i$  to guarantee that ${\cal K}^{(2)}_L$ is real.
Note that the sum in eq. (\ref{here}) runs over all bilinears which can couple to form $A_4$-invariants. Whether
they lead to a non-trivial contribution to the relevant terms depends on the VEVs of the flavons.

For lepton singlets, we can distinguish a diagonal contribution and a non-diagonal one:
\be
{\cal K}_R^{(2)}=[{\cal K}_R^{(2)}]_{d}+[{\cal K}_R^{(2)}]_{nd}
\ee
\be
[{\cal K}_R^{(2)}]_d= \dd\frac{1}{\Lambda_f^2}\sum_{i=1}^5 \left[ (k_e^c)_i (X_i)\bar{e^c}  e^c+(k_\mu^c)_i(X_i)\bar{\mu^c}  \mu^c+(k_\tau^c)_i(X_i) \bar{\tau^c}  \tau^c\right]
\ee
\bea
[{\cal K}_R^{(2)}]_{nd}&=& \sum_{i=2}^5 \left[ \dd\frac{(k_{e\mu}^c)_i}{\Lambda_f^3} (X_i)' \ov{\theta}_{FN}  \bar{e^c}   \mu^c+h.c.\right]\nn\\
\label{here2}
&+& \sum_{i=2}^5 \left[\dd\frac{(k_{e\tau}^c)_i}{\Lambda_f^4} (X_i)'' (\ov{\theta}_{FN})^2  \bar{e^c}  \tau^c  +h.c.\right]\\
&+& \sum_{i=2}^5 \left[\dd\frac{(k_{\mu\tau}^c)_i}{\Lambda_f^3} (X_i)' \ov{\theta}_{FN}  \bar{\mu^c}  \tau^c+h.c.\right] \; .\nn
\eea
Due to the structure of the flavon VEVs only the term with $i=5$ gives a non-vanishing
contribution in the sum in eq. (\ref{here2}). In order to generate a set of supersymmetry breaking soft mass terms, the quantities:
\be
k_I=\{k_0,(k^c_0)_i,(\hat{k}^c_0)_i,k_{u,d}, k_{FN},k_S,k_A,k'_S,k'_A,k_j,(k^c_e)_l,(k^c_\mu)_l,(k^c_\tau)_l,(k^c_{e\mu})_l,(k^c_{e\tau})_l,(k^c_{\mu\tau})_l\}
\ee
are treated as superfields with a constant $\thh\thhb$ component:
\be
k_I=p_I+q_I \thh\thhb m^2_{SUSY}
\label{deck}
\ee
where $p_I$ and $q_I$ do not depend on the Grassmann variables .\footnote{In principle we could allow
for a more general expansion in the Grassmann variables $\theta$ and/or $\bar{\theta}$, including also terms proportional to $\thh$ and to $\thhb$.
These terms can be absorbed in our parametrization after a suitable redefinition of the parameters.}
The quantities $p_I$ and $q_I$ are parameters with absolute values of order one. 
In particular, it is not restrictive to choose (see eq. (\ref{kappa0})):
\be
k_0=1+q_0~ \thh\thhb m_{SUSY}^2~~~,~~~~~~~~~~
(k_0^c)_i=1+(q_0^c)_i~ \thh\thhb m_{SUSY}^2~~~,
\ee
\be
\label{kudexp}
k_{u,d}=1+q_{u,d}~ \thh\thhb m_{SUSY}^2~~~,~~~~~
k_{FN}=1+q_{FN}~ \thh\thhb m_{SUSY}^2~~~.
\ee
When the flavon fields acquire a VEV according to the pattern shown in eqs. (\ref{vevs}) and eq. (\ref{vevtheta}), the K\"ahler potential ${\cal K}$ gives rise to non-canonical kinetic
terms for lepton and slepton  fields of the following form:\footnote
{We adopt  two-component spinor notation, so for example
$e$ ($\bar{e}^c$) denotes the left-handed
(right-handed) component of the electron field.
For instance, in terms of the four-component spinor
$\psi^T_e = (e ~\ov{e}^c)$,
the bilinears $\ov{e} \snub e $ and
$e^c \snu \ov{e}^c$ correspond  to
$\ov{\psi}_{e} \ga^\nu P_L \psi_{e}$ and $\ov{\psi}_e \ga^\nu
P_R \psi_e $ [$P_{L,R} = \frac12 (1\mp \ga^5)$]
respectively.
We  take $\smu \equiv (1,\vec{\sigma})$, $\smub\equiv
(1,-\vec{\sigma})$, $\smn  \equiv \frac14 (\smu \snub -\snu\smub)$,
$\smnb  \equiv \frac14 (\smub \snu -\snub\smu)$ and
$g_{\mu\nu}= {\tt diag}(+1, -1, -1, -1)$, where
$\vec{\sigma} = (\sigma^1, \sigma^2, \sigma^3)$ are the
$2\times 2$ Pauli matrices. Here the four-component matrix
$\gamma^\mu$ is in the chiral basis, where the 2$\times$2 blocks along the diagonal vanish,
the upper-right block is given by $\smu$ and the lower-left block is equal to $\smub$.}
\bea
{\cal L}_{kin}&=&i~ K_{ij}\bar{l}_i \bar{\sigma}^\mu D_\mu l_j+i~ K^c_{ij}\bar{l^c}_i \bar{\sigma}^\mu D_\mu l^c_j\nn\\
&+& K_{ij}\ov{D^\mu \tilde{l}}_i D_\mu \tilde{l}_j+K^c_{ij} \ov{D^\mu \tilde{l}^c}_i  D_\mu \tilde{l}^c_j
\label{kt}
\eea
The matrices $K$ and $K^c$ are given by:
\be
K=
\left(
\begin{array}{ccc}
1+2 t_1~ u& t_4~ u^2& t_5~ u^2\\
\ov{t}_4~ u^2& 1-(t_1+t_2)~ u& t_6~ u^2\\
\ov{t}_5~u^2& \ov{t}_6~ u^2& 1-(t_1-t_2)~ u
\end{array}
\right)~~~,
\label{T}
\ee
\vskip 0.5 cm
\be
K^c=
\left(
\begin{array}{ccc}
1+t^c_1~ u^2+{t'}^c_1~ t^2& t^c_4~ u^2 t& t^c_5~ u^2 t^2\\
\ov{t^c}_4~ u^2 t& 1+t^c_2~ u^2+{t'}^c_2~ t^2& t^c_6~ u^2 t\\
\ov{t^c}_5~ u^2 t^{2}& \ov{t^c}_6~ u^2 t & 1+t^c_3~ u^2+{t'}^c_3~ t^2
\end{array}
\right)~~~.
\label{Tc}
\ee
The coefficients $t_{1,2}$, $t^c_{1,2,3}$ and ${t'}^c_{1,2,3}$ are real, while the remaining coefficients are complex.
As one can see, the corrections to the kinetic terms which render these non-canonical are small, at most at order
$O(u)$ and $O(t^2)$. They have to be taken into account when calculating lepton and slepton masses.
The coefficients $t_i$ and $t^c_i$ are linearly related to the parameters $p_I$ introduced before, see eq. (\ref{deck}). Such a relation
is not particularly significant and in the rest of this paper we will treat $t_i$ and $t^c_i$ as input parameters,
with absolute values of order one. 

Notice that we have neglected possible sub-leading contributions to the kinetic terms of $H_{u,d}$, and to the flavons themselves and $\theta_{FN}$,
which have no impact on the present analysis.
Note further that through the expansion of $k_{u,d}$ as given in eq. (\ref{kudexp}) also the soft masses $m_{H_{u,d}}^2$ for the Higgs doublets
$H_u$ and $H_d$ are generated of size $q_{u,d} m_{SUSY}^2$. Similarly, the expansion of $k_{FN}$ gives rise
to a soft mass term for $\theta_{FN}$ of size $q_{FN} m_{SUSY}^2$.


\subsubsection{Superpotential}
We continue with the discussion of the superpotential:
\be
w=w_l+w_\nu+w_d+w_h+...
\ee
There is a part responsible for the charged lepton masses:
\be
w_l=w_l^{(1)}+w_l^{(2)}+...
\ee
The leading term in the $u$ expansion is:
\be
w_l^{(1)}=\dd\frac{x_e}{\Lambda_f^3} \theta_{FN}^2e^c H_d \left(\varphi_T l\right)
+\dd\frac{x_\mu}{\Lambda_f^2} \theta_{FN}\mu^c H_d \left(\varphi_T l\right)'
+\dd\frac{x_\tau}{\Lambda_f} \tau^c H_d \left(\varphi_T l\right)'' ~~~.
\label{yl1}
\ee
At the next order in $u$ we find:
\be
w_l^{(2)}=\dd\frac{x'_e}{\Lambda_f^4} \theta_{FN}^2e^c H_d \left(\varphi_T^2 l\right)
+\dd\frac{x'_\mu}{\Lambda_f^3} \theta_{FN}\mu^c H_d \left(\varphi_T^2 l\right)'
+\dd\frac{x'_\tau}{\Lambda_f^2} \tau^c H_d \left(\varphi_T^2 l\right)'' ~~~.
\label{yl2}
\ee
To generate both the Yukawa interactions and the soft mass contribution $(m_{e RL}^2)_1$, we regard
the quantities $x_f$ and $x'_f$ as constant superfields, of the type:
\be
x_f=y_f-z_f \thh m_{SUSY}~~~,~~~~~~~x'_f=y'_f-z'_f \thh m_{SUSY}~~~(f=e,\mu,\tau)~~~,
\label{xdecomp}
\ee
where the coefficients $y_f$, $z_f$, $y'_f$ and $z'_f$ have absolute values of order one.
From eqs. (\ref{yl1}) and (\ref{yl2}),
after the breaking of the flavour and the electroweak symmetries, we find the following mass matrix for
the charged leptons:
\begin{equation}
\label{yl_subleading}
m_{l} = \left( \
\begin{array}{ccc}
y_e t^2 u+(y'_e+c'~ y_e) t^2 u^2 & c~ y_e t^2 u^2 & c~ y_e t^2 u^2\\
c~ y_\mu t u^2 & y_\mu t u +(y'_\mu  + c'~ y_\mu) t u^2 & c~ y_\mu t u^2\\
c~ y_\tau u^2 & c~ y_\tau u^2 & y_\tau u+(y'_\tau + c'~ y_\tau) u^2
\end{array}
\right) \frac{v \cos\beta}{\sqrt{2}}
\end{equation}
Note that the matrix $m_l$ is shown in the basis in which the
kinetic terms for the lepton fields are given by eqs. (\ref{kt},\ref{T},\ref{Tc}).
At the LO in the $u$ expansion, $m_l$ matches the matrix shown in eq. (\ref{yf}).
The off-diagonal elements of $m_l$, all proportional to $c$,
originate from the sub-leading contributions to the VEV of the $\varphi_T$ multiplet.
Similarly, the superpotential giving rise to neutrino
masses can be expanded as:
\be
w_\nu=w_\nu^{(1)}+w_\nu^{(2)}+...
\ee
with the LO terms:
\be
w_\nu^{(1)}=
\dd\frac{x_a}{\Lambda_f\Lambda_L} \xi (H_u l H_u l) +\dd\frac{x_b}{\Lambda_f\Lambda_L} (\varphi_S H_u l H_u l)~~~,
\label{here3}
\ee
where $\Lambda_L$ is the scale at which lepton number is violated.
Also in this case $x_{a,b}$ are constant superfields:
\be
x_{a,b}=y_{a,b}+z_{a,b} \thh m_{SUSY}~~~.
\ee
The terms in $w^{(1)}_\nu$
lead to the form of $m_\nu$ as in eq. (\ref{Y}) with $a=y_a c_a$ and $b=y_b c_b$. The next term in the expansion $w_{\nu} ^{(2)}$ is not relevant for our discussion
here, however gives rise to deviations of relative order $u$ from TB mixing.
\noindent The form of $w_d$, responsible for the vacuum alignment of the flavon fields, has already been displayed above in eq. (\ref{wd}).
Finally, the term $w_h$ is associated with the $\mu$ parameter:
\be
w_h=\mu H_u H_d~~~.
\label{muterm}
\ee
This term explicitly breaks the (continuous) $\Uu_{R}$ symmetry of the model, while preserving $R$-parity.
The $\mu$ term might originate from the K\"ahler potential of the theory \cite{giudicemasiero}, after SUSY breaking:
\be
\frac{1}{\Lambda}\int d^2 \theta d^2 \bar\theta (\bar{X} H_u H_d +h.c.)
\ee
where $\Lambda$ is some large scale as, for instance, the Planck scale and $X$ is a chiral superfield, whose $F$ component $F_X$
develops a VEV. This gives rise to  $\mu=\langle F_X\rangle/\Lambda$. In our model we simply assume the existence of the term in eq. (\ref{muterm})
in the superpotential
and allow for an explicit breaking of the $\Uu_{R}$ symmetry controlled by the parameter $\mu$. The soft SUSY breaking term $B\mu$ can then arise from
the $\mu$ term by considering $\mu$ as superfield $\mu + \theta^2 B\mu$. A source of the $\mu$ term within our model are terms which
involve one driving field, a certain number of flavons and the two Higgs doublets $H_u$ and $H_d$. The lowest order term in the
superpotential $w$, allowed by all symmetries of the model, is: 
\begin{equation}
(\varphi^T_0 \varphi_T) H_u H_d/\La_f \, .
\end{equation}
This term generates a contribution to the $\mu$ term of the order of $m_{SUSY} \times u$, when the driving fields acquire VEVs through
the inclusion of soft SUSY breaking terms into the flavon potential \cite{VEVsaux}. The size of such a term is expected to be 
$\lesssim 50$ GeV for $m_{SUSY} \sim O(1 \rm TeV)$.


\subsubsection{Slepton masses}
From the K\"ahler potential and the superpotential we can read off the slepton masses;
they can be parametrized as follows:
\be
-{\cal L}_m \supset
\sum_{i,j=1}^3
\left[
{\bar{\tilde{e}}}_i (m_{eLL}^2)_{ij} \tilde{e}_j +
{\bar{\tilde{e}}}_i (m_{eLR}^2)_{ij} {\bar{\tilde{e^c}}}_j+
{\tilde{e}^c}_i (m_{eRL}^2)_{ij} \tilde{e}_j +
{\tilde{e}^c}_i (m_{eRR}^2)_{ij} {\bar{\tilde{e^c}}}_j
\right]+\sum_{i,j=1}^3
{\bar{\tilde{\nu}}}_i (m_{\nu LL}^2)_{ij} \tilde{\nu}_j
\ee
where $m^2_{(e,\nu)LL}$ and $m^2_{eRR}$ are hermitian matrices and $m^2_{eLR}=(m^2_{eRL})^\dagger$.
In the sneutrino sector only the block $m^2_{\nu LL}$ is present. We neglect contributions to the sneutrino masses
associated to $w_\nu$. Each of these blocks receives several contributions:
\bea
m^2_{(e,\nu)LL}&=&(m^2_{(e,\nu) LL})_K+(m^2_{(e,\nu) LL})_F+(m^2_{(e,\nu) LL})_D~~~,\nn\\
m^2_{eRR}&=&(m^2_{eRR})_K+(m^2_{eRR})_F+(m^2_{eRR})_D+(m^2_{eRR})_{D,FN}~~~.
\eea
The contribution to the slepton masses from the SUSY breaking terms in the K\"ahler potential is given by:
\be
(m^2_{eLL})_K=(m^2_{\nu LL})_K=
\left(
\begin{array}{ccc}
n_0+2 n_1~ u& n_4~ u^2& n_5~ u^2\\
\ov{n}_4~ u^2& n_0-(n_1+n_2)~ u& n_6~ u^2\\
\ov{n}_5~u^2& \ov{n}_6~ u^2& n_0-(n_1-n_2)~ u
\end{array}
\right) m^2_{SUSY}~~~,
\label{m2LL}
\ee
\vskip 0.5 cm
\be
(m^2_{eRR})_K=
\left(
\begin{array}{ccc}
n^c_1& n^c_4~ u^2 t& n^c_5~ u^2 t^2\\
\ov{n^c}_4~ u^2 t&n^c_2& n^c_6~ u^2 t\\
\ov{n^c}_5~ u^2 t^{2}& \ov{n^c}_6~ u^2 t &n^c_3
\end{array}
\right) m^2_{SUSY}~~~,
\label{m2RR}
\ee
where the coefficients are complex, except for $n_{0,1,2}$ and $n^c_{1,2,3}$. \footnote{Additionally, we assume that $n_0$ and $n^c_{1,2,3}$
are positive in order to have positive definite square-masses, to avoid electric-charge breaking minima and 
further sources of electroweak symmetry breaking.}
The coefficients $n_i$ and $n^c_i$ are linearly related to the parameters $q_I$ introduced in eq. (\ref{deck}). Again, such a relation
is not particularly significant and in the rest of this paper we will treat $n_i$ and $n^c_i$ as input parameters,
with absolute values of order one.

The SUSY contribution from the $F$ terms is completely negligible for sneutrinos, i.e. $(m^2_{\nu LL})_F=0$. For charged sleptons
$(m^2_{eLL})_F$ and $(m^2_{eRR})_F$ read:
\be
(m^2_{eLL})_F=m_l^\dagger~(K^c)^{-1}~ m_l ~~~,~~~~~~~(m^2_{eRR})_F=m_l~ (K^{-1})^T~ m_l^\dagger~~~,
\ee
where $m_l$ is the charged lepton mass matrix and $K$, $K^c$ are the matrices specifying the kinetic terms, eqs. (\ref{T}) and (\ref{Tc}).

The SUSY $D$ term contribution is:
\bea
(m^2_{eLL})_D&=&\left(-\frac{1}{2}+\sin^2\theta_W\right) \cos 2\beta ~m_Z^2 K~~~,\nn\\
(m^2_{\nu LL})_D&=&\left(+\frac{1}{2}\right) \cos 2\beta~m_Z^2 K~~~,\\
(m^2_{eRR})_D&=&(-1) \sin^2\theta_W \cos 2\beta~m_Z^2 (K^c)^T~~~\nn,
\eea
where $m_Z$ is the $Z$ mass and $\theta_W$ is the Weinberg angle. Notice again the presence of the matrices $K$ and $K^c$.

For the right-handed charged leptons we find an additional $D$ term contribution stemming from the
fact that $\theta_{FN}$, $e^c$ and $\mu^c$ are charged under the Froggatt-Nielsen symmetry $\Uu_{FN}$, which we assume to be gauged. 
The relevant contribution of the $\Uu_{FN}$ group to the scalar potential (through a $D$ term) is:
\bea
V_{D,FN}&=&\frac{1}{2} \,\left(M_{FI}^2+ g_{FN} \,Q_{FN}^i \dd\frac{\partial {\cal K}}{\partial z_i} z_i\right)^2+ q_{FN} m_{SUSY}^2 \vert \theta_{FN}
\vert^2
\nn\\
&=& g_{FN}^2\, c_\theta m_{SUSY}^2\left( 2|\tilde{e^c}|^2+|\tilde{\mu^c}|^2+t_4^c ~u^2 t~ {\bar{\tilde{e^c}}} \tilde{\mu^c}
+\ov{t^c}_4~ u^2 t~ {\bar{\tilde{\mu^c}}} \tilde{e^c}\right)+...
\eea
where $Q_{FN}^i$ stands for the FN charge of the scalar field $z_i$, and
in the second line we have displayed only the leading contribution to the terms quadratic in the matter fields.
One can check that this contribution is of a similar
form and size as the one originating from the K\"ahler potential, $(m_{e RR}^2)_K$. Thus, we can simply absorb it
by redefining the -anyway- unknown coefficients $n_i^c$, $i=1,...,6$ parametrizing $(m_{e RR}^2)_K$ in eq. (\ref{m2RR}).

Concerning the block $m^2_{eRL}$, this receives three contributions:
\be
m^2_{eRL}=(m^2_{eRL})_1+(m^2_{eRL})_2 + (m^2_{eRL})_3 \; .
\ee
The first one originates from the superpotential, eqs. (\ref{yl1}) and (\ref{yl2}),
and is proportional to the parameters $z_f$ and $z'_f$ of the decomposition in eq. (\ref{xdecomp}):
\begin{equation}
\label{m2RL}
(m^2_{eRL})_1 = A_1 \frac{v \cos\beta}{\sqrt{2}} m_{SUSY}
\end{equation}
with
\be
A_1 =\left( \
\begin{array}{ccc}
z_e t^2 u+(z'_e+c'~ z_e) t^2 u^2 & c~ z_e t^2 u^2 & c~ z_e t^2 u^2\\
c~ z_\mu t u^2 & z_\mu t u +(z'_\mu + c'~ z_\mu) t u^2 & c~ z_\mu t u^2\\
c~ z_\tau u^2 & c~ z_\tau u^2 & z_\tau u+(z'_\tau + c'~ z_\tau) u^2
\end{array}
\right)~~~.
\label{RLNSUSY}
\ee
The second one is related to the fact that the auxiliary fields of the flavon supermultiplet $\varphi_T$ acquire non-vanishing
VEVs of the form as shown in eq. (\ref{VEVsauxphiT}), when soft SUSY breaking terms are included into the flavon potential.
Formally the contribution can be written as: 
\begin{equation}
\overline{\left\langle \frac{\partial w}{\partial \varphi_T} \right\rangle} 
\left\langle \frac{\partial^3 w}{\partial \varphi_T \partial e^c_i \partial l_j} \right\rangle 
\tilde e^c_i \tilde e_j  + h.c.
\end{equation}
so that the soft mass matrix $(m^2_{eRL})_2$ reads:
\begin{equation}
(m^2_{eRL})_2  = A_2 \frac{v \cos\beta}{\sqrt{2}} m_{SUSY}
\end{equation}
with
\be
A_2 = \bar\zeta \, \left( \
\begin{array}{ccc}
y_e t^2 u+(2 y'_e+ \overline{c_F^{\prime}}~ y_e) t^2 u^2 & \overline{c_F}~ y_e t^2 u^2 & \overline{c_F}~ y_e t^2 u^2\\
\overline{c_F}~ y_\mu t u^2 & y_\mu t u +(2 y'_\mu + \overline{c_F^{\prime}}~ y_\mu) t u^2 & \overline{c_F}~ y_\mu t u^2\\
\overline{c_F}~ y_\tau u^2 & \overline{c_F}~ y_\tau u^2 & y_\tau u+(2 y'_\tau + \overline{c_F^{\prime}}~ y_\tau) u^2
\end{array}
\right)~~~.
\label{RLNSUSY2}
\ee
The third contribution is proportional to the charged lepton mass matrix and is enhanced in the large $\tan\beta$ regime:
\be
(m^2_{eRL})_3=-\bar\mu \tan\beta~ m_l~~~.
\label{RLSUSY}
\ee
We remark that the origins of the second and the third contribution are quite similar, since both arise from the auxiliary
component of a superfield: the second one is attributed to the auxiliary component of the flavon superfield $\varphi_T$, 
while the third one originates from the auxiliary component associated to the Higgs doublet $H_d$.

%
%

\section{The physical basis and its stability under renormalization group running}
In this section we first discuss the results for the slepton masses in the physical basis and comment on results
found in the literature. In the second part, we give an estimate of the renormalization group effects on the slepton masses and show in the
leading logarithmic approximation that these effects can be neglected or absorbed into the parametrization of the soft mass terms.

%
%

\subsection{Slepton masses in the physical basis}

All matrices above are given in the basis in which the kinetic terms of the slepton and lepton fields are non-canonical.
To derive the physical masses and the unitary transformations that enter
our computation, we have to go into a basis in which kinetic terms are canonical, for both, slepton and lepton, fields.
Subsequently, we diagonalize the mass matrix of the
charged leptons via a biunitary transformation. To avoid flavour-violating gaugino-lepton-slepton vertices in this intermediate step,
we perform the same transformation
on both fermion and scalar components of the involved chiral superfields.
This procedure, described in detail in appendix B, gives us the physical slepton mass matrices ${\hat m}^2_{(e,\nu)LL}$, ${\hat m}^2_{eRR}$
and ${\hat m}^2_{eRL}$. The results shown here are obtained under the assumption that all parameters of the model are real.
The analytical expressions for the slepton mass matrices in the physical basis contain the first non-vanishing order in each of the matrix elements.
We start with the left-left (LL) block. The contribution from the soft breaking terms is common to charged sleptons and sneutrinos and reads:
\begin{equation}
\label{eq:m_LL_hat}
\begin{array}{ll}
(\hat{m}_{eLL}^2)_K &= (\hat{m}_{\nu LL}^2)_K\nn\\

&\!\!\!\!\!\!\!\!\!\!\!\!\!\!\!\!=\left( \begin{array}{ccc}
                n_0 + 2 \, \hat{n}_1 \, u
                                & (\hat{n}_4 + (3 \, \hat{n}_1 + \hat{n}_2) \, c) \, u^2
                                & (\hat{n}_5 + (3 \, \hat{n}_1 - \hat{n}_2) \, c) \, u^2\\
                 (\hat{n}_4 + (3 \, \hat{n}_1 + \hat{n}_2) \, c) \, u^2
                                & n_0 - (\hat{n}_1 + \hat{n}_2) \, u
                                & (\hat{n}_6 - 2 \, \hat{n}_2 \, c) \, u^2 \\
                (\hat{n}_5 + (3 \, \hat{n}_1 - \hat{n}_2) \, c) \, u^2
                                & (\hat{n}_6 - 2 \, \hat{n}_2 \, c) \, u^2
                                & n_0 - (\hat{n}_1 - \hat{n}_2) \, u

\end{array}
\right) \, m_{SUSY}^2
\end{array}
\end{equation}
\begin{equation}
\label{eq:m_LL_hat_coefficients}
\hat{n}_i = n_i - t_i n_0 \;\;\; \mbox{for} \;\;\; i=1,2,4,5,6\\
\end{equation}
The supersymmetric $F$ and $D$ term contributions are given by:
\begin{equation}
(\hat{m}_{eLL}^2)_F=\hat{m}_l^T \hat{m}_l~~~,~~~~~~~(\hat{m}_{\nu LL}^2)_F=0
\end{equation}
and
\begin{equation}
(\hat{m}^2_{eLL})_D=\left(-\frac{1}{2}+\sin^2\theta_W \right) \cos 2\beta ~m_Z^2 \times \mathbb{1}~~~,~~~~~~~
(\hat{m}^2_{\nu LL})_D=\left(+\frac{1}{2} \right) \cos 2\beta~m_Z^2 \times \mathbb{1}~~~,
\end{equation}
with $\hat{m}_l$ being the mass matrix for the charged leptons in the same basis, i.e. diagonal and with
canonically normalized kinetic terms. The supersymmetric $D$ term contributions are proportional to the unit matrix.
Notice that in the physical basis all SUSY contributions are diagonal in flavour space.
Both, the $F$ and the $D$ term, contributions are small compared to that
coming from the K\"ahler potential. The relative suppression is of order $\hat{m}_l^T \hat{m}_l/m_{SUSY}^2$
and $m_Z^2/m_{SUSY}^2$, respectively, which do not exceed the per cent level for typical values of $m_{SUSY}$ around 1 TeV.
Note also that the SUSY part is the only one that distinguishes between charged sleptons
and sneutrinos.
The dominant mass matrix, $(\hat{m}_{eLL}^2)_K=(\hat{m}_{\nu LL}^2)_K$, has a structure which is very similar
to that of the corresponding matrix in the original basis, i.e.
all matrix elements in the two bases are of the same order in the $(u,t)$ expansion. They only differ
for coefficients of order one.

For $\hat{m}_{eRR}^2$ we find that $(\hat{m}_{eRR}^2)_K$ is given by:
\begin{equation}
\label{eq:m_RR_hat}
(\hat{m}_{eRR}^2)_K = \left( \begin{array}{ccc}
                 n_1^c
                                & 2 \, c \,  (n_1^c - n_2^c) \, \dd\frac{m_e}{m_\mu} u
                                & 2 \, c \, (n_1^c - n_3^c) \, \dd\frac{m_e}{m_\tau} u\\[0.1in]
                2 \, c  \, (n_1^c - n_2^c) \, \dd\frac{m_e}{m_\mu} u
                                & n_2^c
                                & 2 \, c \, (n_2^c - n_3^c) \, \dd\frac{m_\mu}{m_\tau} u\\[0.1in]
                2 \, c  \, (n_1^c - n_3^c) \, \dd\frac{m_e}{m_\tau} u
                                & 2 \, c  \, (n_2^c - n_3^c) \,\dd \frac{m_\mu}{m_\tau} u
                                & n_3^c
\end{array}
\right) \, m_{SUSY}^2~~~.
\end{equation}
The supersymmetric terms are:
\begin{equation}
(\hat{m}_{eRR}^2)_F=\hat{m}_l \hat{m}_l^T~~~\mbox{and}~~~
(\hat{m}^2_{eRR})_D=-\sin^2\theta_W \cos 2\beta ~m_Z^2 \times \mathbb{1}~~~.
\end{equation}
Also in this case the SUSY contributions are diagonal and numerically negligible in
most of our parameter space. 
The dominant contribution is thus $(\hat{m}_{eRR}^2)_K$.
We note that $(\hat{m}_{eRR}^2)_K$ at variance with $(\hat{m}_{(e,\nu) LL}^2)_K$ does not depend on the parameters describing the non-canonical
kinetic terms.
Comparing the size of the entries of $(\hat{m}_{eRR}^2)_K$ with those of $(m_{eRR}^2)_K$
we see that the diagonal elements are still of the same order in $t$ and $u$, whereas all off-diagonal
elements are enhanced by a factor $1/u$. This can be understood in terms of the rotation done on the right-handed leptons to
diagonalize the charged lepton mass matrix. Such a rotation is characterized by small angles
proportional to $c~ y_e/y_\mu~ t u$, $ c~ y_e/y_\tau~ t^2 u$ and $c~ y_\mu/y_\tau~ t u$ in the sectors 12, 13 and 23,
respectively. By making the same rotation on the corresponding right-handed sleptons,
we obtain the off-diagonal terms of $\hat{m}_{eRR}^2$. 

Finally, coming to the RL block of the mass matrix for charged sleptons, we find:
\begin{equation}
\label{eq:m_RL_hat}
(\hat{m}_{eRL}^2)_{1} = \left( \begin{array}{ccc}
                \dd\frac{z_e}{y_e} \, m_e
                                & 2 c  \, \dd\frac{(z_e y_\mu - z_\mu y_e)} {y_e y_\mu}\, m_e u
                                & 2 c\, \dd\frac{(z_e y_\tau - z_\tau y_e)}{y_e y_\tau}  \, m_e u\\[0.1in]
                c \, \dd\frac{(z_\mu y_\mu^\prime - z_\mu^\prime y_\mu)}{y_\mu^2} \, m_\mu u^2
                                & \dd\frac{z_\mu}{y_\mu} m_\mu
                                & 2 c  \,\dd\frac{(z_\mu y_\tau -  z_\tau y_\mu)}{y_\mu y_\tau} m_\mu u\\[0.1in]
                c  \, \dd\frac{(z_\tau y_\tau ^\prime - z_\tau ^\prime y_\tau)}{y_\tau^2} \, m_\tau u^2
                                & c \, \dd\frac{(z_\tau y_\tau ^\prime - z_\tau ^\prime y_\tau)}{y_\tau^2} \, m_\tau u^2
                                & \dd\frac{z_\tau}{y_\tau} \, m_\tau
\end{array}
\right) \, m_{SUSY} \; ,
\end{equation}
\begin{equation}
\label{eq:m_RL_hat_2}
(\hat{m}_{eRL}^2)_{2} = \zeta \left( \begin{array}{ccc}
                m_e
                                & (c_F -c) \, m_e u
                                & (c_F -c) \, m_e u\\[0.1in]
               (c_F -c) \, m_\mu u
                                & m_\mu
                                & (c_F -c) \, m_\mu u\\[0.1in]
               (c_F -c) \, m_\tau u
                                & (c_F -c) \, m_\tau u
                                & m_\tau
\end{array}
\right) \, m_{SUSY} \; ,
\end{equation}
and
\be
\label{eq:m_RL_hat_3}
(\hat{m}^2_{eRL})_3=-\mu \tan\beta~ \hat{m}_l~~~.
\ee
The matrix $\hat{m}_{eRL}^2$, which is the sum of these three contributions, does not depend on the parameters describing the non-canonical
kinetic terms through $K$ and $K^c$. 
An important feature of $(\hat{m}_{eRL}^2)_1$ is that the elements below the diagonal
are suppressed by a factor $u$ compared to the corresponding elements
of $(m_{eRL}^2)_1$, i.e. before the transformations for canonical normalization
of the kinetic terms and diagonalization of the charged lepton mass matrix are applied.
However, this does not happen for the second contribution associated to the non-vanishing VEVs of the auxiliary components of the supermultiplet $\varphi_T$
so that the elements of $(\hat{m}_{eRL}^2)_{2}$ are still of the same order in the expansion parameters $t$ and $u$ as those of the
matrix $(m_{eRL}^2)_{2}$. Nevertheless there are cases in which this contribution can be suppressed. In the first case the
VEVs of the auxiliary fields contained in the supermultiplet $\varphi_T$ vanish, i.e. the parameter $\zeta$ is zero, due to the fact that the soft SUSY breaking
terms in the flavon potential are (assumed to be) universal, that is equal to the terms of the superpotential $w_d$ up to an
overall proportionality constant \cite{VEVsaux}. The second possibility arises, if the VEVs of the auxiliary fields
can be completely aligned with those of the flavon $\varphi_T$ at LO as well as NLO, such that $c_F$ becomes equal to $c$.
In both cases the off-diagonal elements of $(\hat{m}_{eRL}^2)_{2}$ are further suppressed than shown in eq. (\ref{eq:m_RL_hat_2}).
We emphasize this fact here, since it turns out that the suppression of the off-diagonal elements below the diagonal as it occurs in the
case of $(\hat{m}_{eRL}^2)_{1}$ is relevant for the actual size of the leading behaviour of the normalized branching ratios
$R_{ij}$ with respect to the expansion in $u$. As we shall see in section 4,
in a general case $R_{ij} \propto u^2$ holds, whereas, if the contribution
in eq. (\ref{eq:m_RL_hat_2}) vanishes or is also suppressed, $R_{ij}$ is proportional to $u^4$.
The contribution $(\hat{m}^2_{eRL})_3$ is diagonal in flavour space.
Concerning the possible size of this contribution, note that $|\mu| \tan \beta/m_{SUSY}$ is
the relative magnitude of the non-vanishing elements of $(\hat{m}_{eRL}^2)_3$ with respect to the corresponding
ones in $(\hat{m}_{eRL}^2)_{1,2}$. 
Notice finally that the (31) and (32) element of $\hat{m}_{eRL}^2$ coincide.

We can compare our results with those found in \cite{jap}, where the slepton mass matrices for a model possessing the same flavour symmetry
were also estimated in a similar framework. The main difference between the two setups is that in our model
SUSY is a softly broken global symmetry, whereas in \cite{jap} the model has been embedded into SUGRA.
We agree on the structure of the matrix $(\hat{m}_{eLL}^2)_K$ ($\tilde{m}^2_L$ in the notation of \cite{jap}).
Concerning the matrix $(\hat{m}_{eRR}^2)_K$
($\tilde{m}^2_R$ in \cite{jap}), we see that the off-diagonal matrix elements of $\tilde{m}^2_R$ are all of order $u^2$, whereas we find that those of $(\hat{m}_{eRR}^2)_K$ are of order $u$.
Such a discrepancy has a minor impact on the estimate of the rates for the radiative transitions, since, as we shall see in the next section,
the RR block gives a subdominant contribution. It is interesting to note that also in the SUGRA context analyzed
in \cite{jap} the VEVs of the auxiliary components of the flavon supermultiplets give rise to $(\tilde{m}^2_{LR})_{21}\approx m_\mu m_{SUSY}~ u$,
which corresponds to our $(\hat{m}_{eRL}^2)_{2,21}\approx  m_\mu m_{SUSY} ~u$,
with similar implication on the rate of $\mu\to e \gamma$.
%
%

\subsection{Estimate of renormalization group effects}

Here we briefly comment on possible effects of the running from the scale $\Lambda_f\approx \Lambda_L$
at which the sfermion masses originate in our effective theory, down to the low energy scale, at which the amplitudes of LFV transitions are
evaluated.
The renormalization group equations for
the soft mass terms $(\hat{m}^2_{(e,\nu)LL} )_{K}$,  $(\hat{m}^2_{eRR} )_{K}$ and
$\hat{A}_e \equiv \sqrt{2}[(\hat{m}^2_{eRL})_{1} + (\hat{m}^2_{eRL})_{2}]/(v \cos \beta)$, 
denoting $ t' \equiv \log (\Lambda_L/m_{SUSY})$, are \cite{RGEs}:

\begin{eqnarray}
16\pi^2 \frac{d}{d t'} \left( \hat{m}^2_{eLL} \right)_{Kij}
&=&   -\left( \frac{6}{5} g_1^2 \left| M_1 \right|^2
+ 6 g_2^2 \left| M_2 \right|^2 \right) \delta_{ij}
-\frac{3}{5} g_1^2~S~\delta_{ij} \nonumber \\
&+&  \left (( \hat{m}^2_{eLL} )_{K} \hat{Y}_e^{\dagger} \hat{Y}_e
+ \hat{Y}_e^{\dagger} \hat{Y}_e ( \hat{m}^2_{eLL} )_{K} \right)_{ij} \nonumber \\
&+& 2 \left( \hat{Y}_e^{\dagger} (\hat{m}^2_{eRR} )_{K} \hat{Y}_e
           +{m}^2_{H_d} \hat{Y}_e^{\dagger} \hat{Y}_e
+\hat{A}_e^{\dagger} \hat{A}_e \right)_{ij} \nonumber~~~, \\
16\pi^2 \frac{d}{d t'} \left( \hat{m}^2_{eRR} \right)_{Kij} &=&
- \frac{24}{5} g_1^2 \left| M_1 \right|^2 \delta_{ij} + \frac{6}{5} g_1^2~S~\delta_{ij} \nonumber \\
&+& 2 \left ( ( \hat{m}^2_{eRR} )_{K}  \hat{Y}_e \hat{Y}_e^{\dagger} + \hat{Y}_e \hat{Y}_e^{\dagger} ( \hat{m}^2_{eRR} )_{K} \right)_{ij}
\nonumber \\
\label{RGE}
&+& 4 \left( \hat{Y}_e ( \hat{m}^2_{eLL} )_{K} \hat{Y}_e^{\dagger} + {m}^2_{H_d}
\hat{Y}_e \hat{Y}_e^{\dagger} +  \hat{A}_e \hat{A}_e^{\dagger} \right)_{ij}~~~, \\
16\pi^2 \frac{d}{d t'}  \left( \hat{A}_e \right)_{ij} &=&
 \left( -\frac{9}{5} g_1^2 -3 g_2^2
+ 3 {\rm Tr} ( \hat{Y}_d^{\dagger} \hat{Y}_d )
+   {\rm Tr} ( \hat{Y}_e^{\dagger} \hat{Y}_e ) \right )  \left(\hat{A}_{e}\right)_{ij} \nonumber \\
&+& 2 \left(
\frac{9}{5} g_1^2 M_1 + 3 g_2^2 M_2
+ 3 {\rm Tr} ( \hat{Y}_d^{\dagger} \hat{A}_d)
+   {\rm Tr} ( \hat{Y}_e^{\dagger} \hat{A}_e) \right) \hat{Y}_{e_{ij}} \nonumber \\
&+& 4 \left( \hat{Y}_e \hat{Y}_e^{\dagger} \hat{A}_e \right)_{ij}
+ 5 \left(\hat{A}_e \hat{Y}_e^{\dagger} \hat{Y}_e \right)_{ij}~~~, \nonumber \\
16\pi^2 \frac{d}{d t'} \hat{Y}_{e_{ij}} &=& \left ( -\frac{9}{5} g_1^2 - 3 g_2^2
  + 3 \,{\rm Tr} ( \hat{Y}_d \hat{Y}_d^{\dagger})
  +   {\rm Tr} ( \hat{Y}_e \hat{Y}_e^{\dagger})
\right ) \hat{Y}_{e_{ij}}\nonumber \\
&+& 3 \, \left( \hat{Y}_e \hat{Y}_e^{\dagger} \hat{Y}_e \right)_{ij} \nonumber~~~,
\end{eqnarray}

where $g_{1,2}$ are the gauge couplings \footnote{In the GUT normalization, such that $g_2=g$ and $g_1=\sqrt{5/3} g'$.} of SU(2)$_L\times \Uu_Y$, $M_{1,2}$ the corresponding gaugino mass terms, $\hat{Y}_{e,d}\equiv \sqrt{2} \hat{m}_{l,d}/(v \cos\beta)$ are the Yukawa matrices for charged leptons and down quarks, $\hat{A}_d=\sqrt{2}[(\hat{m}^2_{dRL})_{1} + (\hat{m}^2_{dRL})_{2}]/(v \cos \beta)$ and:
\begin{equation}
S = {\rm Tr} (\hat{m}^2_{qLL} + \hat{m}^2_{dRR}- 2 \hat{m}^2_{uRR}
- (\hat{m}^2_{eLL})_K + (\hat{m}^2_{eRR})_K ) - {m}^2_{H_d}
+ {m}^2_{H_u} \nonumber.
\end{equation}
The matrix $(\hat{m}^2_{\nu LL})_K$ coincides with $(\hat{m}^2_{e LL})_K$ and has the same evolution.
For squarks we have introduced soft mass terms analogous to those previously discussed for sleptons. 
To estimate the corrections to the slepton masses induced by the renormalization group evolution we adopt the leading logarithmic approximation and
substitute each of the running quantities with their initial conditions at the scale $\Lambda_L\approx\Lambda_f$ in eqs. (\ref{RGE}). In particular, for the matrices $(\hat{m}^2_{(e,\nu)LL})_K$, $(\hat{m}^2_{eRR})_K$ and
$\hat{A}_e$, these initial conditions are given in eqs.  (\ref{eq:m_LL_hat}), (\ref{eq:m_RR_hat}) and (\ref{eq:m_RL_hat}), 
(\ref{eq:m_RL_hat_2}), respectively. In this approximation one easily sees that
the largest corrections to the matrices $(\hat{m}^2_{(e,\nu)LL})_K$ and $(\hat{m}^2_{eRR})_K$ come from electroweak gauge interactions and are proportional to the identity matrix in flavour space. Due to the negative sign of the dominant contribution these diagonal elements
increase by evolving the mass matrices from the cutoff scale down to the electroweak scale. 
This effect is taken into account in our numerical study presented in
section 5. \footnote{This effect can also be viewed as a redefinition of the initial parameters $n_0$ and $n_{1,2,3}^c$.} Each off-diagonal element of
$(\hat{m}^2_{(e,\nu)LL})_K$ receives at most a relative correction of order:
\be
\dd\frac{1}{16 \pi^2}~ u~ \log\left(\dd\frac{\Lambda_L}{m_{SUSY}}\right)~~~,
\ee
while those to $(\hat{m}^2_{eRR})_K$ are even more suppressed, i.e.
\be
\dd\frac{1}{16 \pi^2}~ u^2~ \log\left(\dd\frac{\Lambda_L}{m_{SUSY}}\right)~~~.
\ee
All such contributions can be safely neglected.

The matrix $\hat{A}_e$ gets a first correction by an overall multiplicative factor that can be absorbed, for instance, by a common
rescaling of the parameters, and a second correction of the type $\hat{A}_e\to\hat{A}_e+ K~ \hat{Y}_e$, which can be absorbed by the
redefinition $z_f^{(\prime)}\to z_f^{(\prime)}+k~ y_f^{(\prime)}$, where $K$ and $k$ are constants. In our numerical study these effects are treated in this way. Finally, additional corrections to the off-diagonal elements of $\hat{A}_e$ are negligible.
We can conclude that the corrections induced by the RG running are either negligible or could be absorbed in our parametrization.
Thus, in our model the soft mass terms of sleptons are completely controlled by the flavour symmetry and by its spontaneous breaking.
We recall that our model does not contain right-handed neutrinos and ignores the dynamics above the scale $\Lambda_L\approx\Lambda_f$. 
Notice that the Yukawa couplings of the charged leptons, $\hat{Y}_e$,
remain diagonal during the evolution.

In a see-saw version,
one should also include the effects of the running from the cutoff scale down to the right-handed neutrino mass scale(s)
and the corresponding threshold effects.
In this case the previous conclusions might change, since we expect for generic order one contributions from RG running that they
enter the amplitudes of the branching ratios at the level of $1/(16 \pi^2)$, due to the loop suppression, which is (roughly) equal
to a contribution of order $u^2$ in the amplitude in our context.


\section{Results in the mass insertion approximation}
We can now evaluate the normalized branching ratios $R_{ij}$ for the LFV transitions
$\mu\to e \gamma$, $\tau\to\mu\gamma$ and $\tau\to e \gamma$. 
In this section we establish the leading dependence of the 
quantities $R_{ij}$ on the symmetry breaking parameter $u$.
We recall that in the class of models based on the flavour symmetry group $A_4\times Z_3\times U(1)_{FN}$
an estimate based on an effective Lagrangian approach suggests that
$R_{ij}$ generically scales as $u^2$, which can be reconciled with the present bound on $R_{\mu e}$
only if the scale $M$ of new physics is sufficiently large, above 10 TeV.
The effective Lagrangian approach also indicates that, under certain conditions, in the 
SUSY case a cancellation might take place, and $R_{ij}$ might scale as a combination of two terms,
one proportional to $u^4$ and one proportional to $(m_j^2/m_i^2) u^2$, as shown in eq. (\ref{LFVsusy}),
thus allowing for a substantially smaller scale of new physics, in the range of $(1\div10)$ TeV. 
When $R_{\mu e}$ is dominated by a one-loop amplitude with virtual particles of mass $m_{SUSY}$, $M$ 
and $m_{SUSY}$ are roughly related by $M=(4\pi/g) m_{SUSY}$ and a given lower bound on $M$
corresponds to a lower bound on $m_{SUSY}$ one order of magnitude smaller.
 Here we determine
the actual leading order behaviour of $R_{ij}$ by expressing the result as a 
power series in the parameter $u$.
\subsection{Analytic results}
In order to do so it is useful to first analyse the predictions in the so-called mass insertion (MI)
approximation, where we have a full control of the results in its analytic form.
A more complete discussion based on one-loop results can be found in section 5.
For the case at hand, the MI approximation consists in expanding the amplitudes in powers of the off-diagonal elements
of the slepton mass matrices, normalized to their average mass. From the expression of the mass matrices of the previous section we see that in our case such an expansion
amounts to an expansion in the parameters $u$ and $t$, which we can directly compare with eq. (\ref{LFVsusy}).
A common value in the diagonal entries of both
LL and  RR blocks is assumed and we consequently set $n_0=n_1^c=n_2^c=n_3^c=1$ and also $\hat{n}_1=\hat{n}_2=0$
in this section, so that the average mass becomes $m_{SUSY}$. On the contrary, no assumptions have been made for the trilinear soft terms, which keep the expression as in eqs. (\ref{eq:m_RL_hat}-\ref{eq:m_RL_hat_3}). Concerning chargino and neutralino mass matrices, they carry a dependence
on the vector boson masses $m_{W,Z}$ through off-diagonal matrix elements.
Such a dependence is not neglected in this approximation, but only the leading order term of an expansion
in $m_{W,Z}$ over the relevant SUSY mass combination is kept. At the same time, to be consistent, we have to neglect the supersymmetric
contributions of $\hat{m}^2_{\nu LL}$ and $\hat{m}^2_{e LL}$ and therefore $\hat{m}^2_{\nu LL}$ and $\hat{m}^2_{e LL}$ coincide.
Using these simplifications, the ratios $R_{ij}$ can be expressed as:
\be
R_{ij}= \frac{48\pi^3 \alpha}{G_F^2 m_{SUSY}^4}
\left(\vert A_L^{ij} \vert^2+\vert A_R^{ij} \vert^2 \right)~~~.
\label{rij}
\ee
At the LO, the amplitudes $A_L^{ij}$ and $A_R^{ij}$ are given by:
\bea
A_L^{ij}&=&a_{LL} (\delta_{ij})_{LL} + a_{RL} \frac{m_{SUSY}}{m_i} (\delta_{ij})_{RL}\nn\\
A_R^{ij}&=&a_{RR} (\delta_{ij})_{RR} + a_{LR} \frac{m_{SUSY}}{m_i} (\delta_{ij})_{LR}
\label{ALAR}
\eea
where $a_{CC'}$ $(C,C'=L,R)$ are dimensionless functions of the ratios
$M_{1,2}/m_{SUSY}$, $\mu/m_{SUSY}$ and of $\tan\theta_W$. Their typical size is one tenth of $g^2/(16\pi^2)$,
$g$ being the SU(2)$_L$ gauge coupling constant. In our conventions their explicit expression is given by:
\bea
a_{LL}&=&\dd\frac{g^2}{16\pi^2}\left[ f_{1n}(a_2)+f_{1c}(a_2)+
\dd\frac{M_2\mu\tan\beta}{M_2^2-\mu^2}\Big(f_{2n}(a_2,b)+f_{2c}(a_2,b)\Big)\right.\nn\\
&+&\left.\tan^2\theta_W\left(f_{1n}(a_1)- \dd\frac{M_1\mu\tan\beta}{M_1^2-\mu^2}f_{2n}(a_1,b)- M_1\left(\left(\dd\frac{z_i}{y_i}+\zeta\right) m_{SUSY}-\mu\tan\beta\right)\dfrac{f_{3n}(a_1)}{m_{SUSY}^2}\right)\right]\nn\\
a_{RL}&=&\dd\frac{g^2}{16\pi^2}\tan^2\theta_W\dd\frac{M_1}{m_{SUSY}} 2 f_{2n}(a_1)\\
a_{RR}&=&\dd\frac{g^2}{16\pi^2}\tan^2\theta_W \left[4 f_{1n}(a_1)+ 2\dd\frac{M_1\mu\tan\beta}{M_1^2-\mu^2}f_{2n}(a_1,b)- M_1\left(\left(\dd\frac{z_i}{y_i}+\zeta\right) m_{SUSY}-\mu\tan\beta\right)\dfrac{f_{3n}(a_1)}{m_{SUSY}^2}\right]\nn\\
a_{LR}&=&\dd\frac{g^2}{16\pi^2}\tan^2\theta_W\dd\frac{M_1}{m_{SUSY}} 2 f_{2n}(a_1)\nn
\label{MIcoefficients}
\eea
where $a_{1,2}=M^2_{1,2}/m_{SUSY}^2$, $b=\mu^2/m_{SUSY}^2$ and $f_{i(c,n)}(x,y)=f_{i(c,n)}(x)-f_{i(c,n)}(y)$.
The functions $f_{in}(x)$ and $f_{ic}(x)$, slightly different from those in \cite{Ciuchini:2007ha}, are given by:
\bea
f_{1n}(x)&=&(-17 x^3+9 x^2+9 x-1+6 x^2(x+3) \log x)/(24(1-x)^5)\nn\\
f_{2n}(x)&=&(-5 x^2+4 x+1+2x(x+2)\log x)/(4(1-x)^4)\nn\\
f_{3n}(x)&=&(1+9x -9x^2-x^3+6x(x+1) \log x)/(2(1-x)^5)\\
f_{1c}(x)&=&(-x^3-9x^2+9x+1+6x(x+1) \log x)/(6(1-x)^5)\nn\\
f_{2c}(x)&=&(-x^2-4 x+5+2(2x+1)\log x)/(2(1-x)^4)~~~.\nn
\label{MIfunctions}
\eea
\begin{table}[!ht]
\centering
    \begin{math}
    \begin{array}{|c|c|c|c|c|}
        \hline
        &&&& \\[-9pt]
       ij &w^{LL}_{ij}  &w^{RL}_{ij} &w^{RR}_{ij} &w^{LR}_{ij}\\[10pt]
        \hline
        &&&&\\[-9pt]
        \mu e &\hat{n}_4 &\zeta (c_F-c)& 0 &2   \, \dd\frac{(z_e y_\mu - z_\mu y_e)} {y_e y_\mu}\,~c +\zeta (c_F-c)\\[3pt]
        \hline
        &&&&\\[-9pt]
        \tau e &\hat{n}_5 &\zeta (c_F-c)&  0 &2 \, \dd\frac{(z_e y_\tau - z_\tau y_e)}{y_e y_\tau} ~c+\zeta (c_F-c)\ \\[3pt]
        \hline
        &&&&\\[-9pt]
        \tau \mu &\hat{n}_6 &\zeta (c_F-c)& 0 &2  \,\dd\frac{(z_\mu y_\tau -  z_\tau y_\mu)}{y_\mu y_\tau}~c+\zeta (c_F-c) \\[3pt]
        \hline
    \end{array}
    \end{math}
\caption{Coefficients $w^{CC'}_{ij}$ characterizing the transition amplitudes for $\mu\to e \gamma$, $\tau\to e \gamma$ and $\tau\to \mu\gamma$, in the MI approximation in which $n_0$ and $n_i^c$ are set to one and $\hat{n}_{1,2}$ to zero so that $w^{RR}_{ij}$ vanish.}
\label{table_coefficients}
\end{table}
Notice that $a_{CC'}$ do neither depend on $u$ nor on the fermion masses $m_{i,j}$.
Finally, $(\delta_{ij})_{CC'}$ parametrize the MIs and are defined as:
\be
(\delta_{ij})_{CC'}=\frac{(\hat{m}^2_{eCC'})_{ij}}{m^2_{SUSY}}~~~.
\ee
From the mass matrices of the previous section, we find ($j<i$):
\be
\begin{array}{ll}
(\delta_{ij})_{LL}=w^{LL}_{ij} u^2~~~,&
(\delta_{ij})_{RL}=\dd\frac{m_i}{m_{SUSY}} \left( w^{RL}_{ij} u +w^{'RL}_{ij}  u^2\right)\\[12pt]
(\delta_{ij})_{RR}=w^{RR}_{ij} \dd\frac{m_j}{m_i} u~~~,&
(\delta_{ij})_{LR}=w^{LR}_{ij} \dd\frac{m_j}{m_{SUSY}} u~~~.
\end{array}
\label{deltas}
\ee
where for the mass insertion $(\delta_{ij})_{RL}$ we have also displayed the NLO contributions, in order to better compare
our results with those of the effective Lagrangian approach.
The explicit expression for the LO coefficients $w^{CC'}_{ij}$ are listed in table 2. Also the NLO coefficients
$w^{'RL}_{ij} $ are dimensionless combinations of order one parameters.
By substituting the mass insertions of eq. (\ref{deltas}) into the amplitudes $A_{L,R}^{ij}$ of eq. (\ref{ALAR})
and by using eq. (\ref{rij}), we get:
\be
R^{SUSY}_{ij}= \frac{48\pi^3 \alpha}{G_F^2 M^4}\left[\vert w^{(0)}_{ij} u\vert^2+ 2 w^{(0)}_{ij} w^{(1)}_{ij} u^3+
\vert w^{(1)}_{ij} u^2\vert^2+\frac{m_j^2}{m_i^2} \vert w^{(2)}_{ij} u\vert^2\right]
\label{RSUSY}
\ee
with $M= (4 \pi/g) m_{SUSY}$ and 
\bea
w^{(0)}_{ij}&=&\dd\frac{16 \pi^2}{g^2} a_{RL} w^{RL}_{ij}~~,\nn\\
w^{(1)}_{ij}&=&\dd\frac{16 \pi^2}{g^2} \left(a_{LL} w^{LL}_{ij}+a_{RL}  w^{'RL}_{ij}\right)~~,\nn\\
w^{(2)}_{ij}&=&\dd\frac{16 \pi^2}{g^2} \left(a_{RR} w^{RR}_{ij}+a_{LR} w^{LR}_{ij}\right)~~~.
\label{wcoeff}
\eea
The behaviour displayed in eq. (\ref{RSUSY}) differs from the one expected on the basis of the effective Lagrangian approach in the SUSY case, eq. (\ref{LFVsusy}).
This is due to the presence of the term $w^{(0)}_{ij} \propto w^{RL}_{ij}$. Assuming $w^{RL}_{ij}=0$ we recover what is expected from the effective Lagrangian approach
in the SUSY case, whereas when $w^{RL}_{ij}$ does not vanish, the LO behaviour matches the prediction of the effective Lagrangian 
approach in the generic, non-supersymmetric case, eq. (\ref{LFV}). As shown in table 2, the coefficient $w^{RL}_{ij}$ is universal, namely it is independent
from the flavour indices and it vanishes in two cases: 
\begin{itemize}
\item[i)]
$c_F=c$, which reflects the alignment of the VEVs of the scalar
and auxiliary components of the flavon supermultiplet $\varphi_T$, see eqs. (\ref{vevs}) and (\ref{VEVsauxphiT}).
\item[ii)]
$\zeta=0$ which can be realized by special choices of the soft SUSY breaking terms
in the flavon sector, i.e. the assumption of universal soft SUSY breaking terms in the flavon potential. 
\end{itemize}
In our model none of these possibilities is natural, see \cite{VEVsaux}, and both require a tuning of the underlying parameters.
If $w^{RL}_{ij}=0$, the result expected from the effective Lagrangian approach in the SUSY case is obtained in a non-trivial way. Indeed, it is a consequence of a cancellation taking place
when going from the Lagrangian to the physical basis. In particular, for $w^{RL}_{ij}=0$ the MI $(\delta_{ij})_{RL}$ scales as  $m_i u^2/m_{SUSY}$
and not as $m_i u/m_{SUSY}$ as  we might naively guess by looking at eq. (\ref{RLNSUSY}).
As a consequence $R^{SUSY}_{ij}$ scales as $u^4$ and not as $u^2$
for $m_j=0$. 

In the general case when $w^{RL}_{ij}$ is non-vanishing, the dominant contribution to $R_{ij}^{SUSY}$ regarding the expansion in $u$
is flavour independent and, at the LO in the $u$ expansion,
we predict $R_{\mu e}=R_{\tau\mu}=R_{\tau e}$, at variance with the predictions
of most of the other models, where, for instance, $R_{\mu e}/R_{\tau \mu}$ can be much smaller than one \cite{Raidal,MLFV,SUSYLFV+symmetries}.
If $w^{RL}_{ij}$ is non-vanishing, it is interesting to analyze the relative weight of the leading and sub-leading contributions to $R_{ij}$. 
For this purpose we list in table \ref{table_aCC} the expressions and the numerical values of the functions $a_{CC'}$, in the
limit $\mu=M_{1,2}=m_{SUSY}$. 
\begin{table}[!ht]
\hspace{-0.4in}    \begin{math}
    \begin{array}{|c|l|c|}
        \hline
      &&  \\[3pt]
        a_{LL}&\dd\frac{1}{240}\frac{g^2}{16 \pi^2}\left[1-3\left(1+4\left(\dfrac{z_i}{y_i}+\zeta\right)\right)\tan^2\theta_W+4 \Big(4+5\tan^2\theta_W\Big)\tan\beta\right]&+(2.0\div16.3) \\
 &&   \\[3pt]
        \hline
     &&   \\[3pt]
        a_{RL}=a_{LR}&\dd\frac{1}{12}\frac{g^2}{16 \pi^2}\tan^2\theta_W&0.30 \\
    &&    \\[3pt]
        \hline
  &&      \\[3pt]
        a_{RR}&\dd\frac{1}{60}\frac{g^2}{16 \pi^2}\tan^2\theta_W\left[-3-3\left(\dfrac{z_i}{y_i}+\zeta\right)-\tan\beta\right]&-(0.5\div1.3) \\
    &&     \\[3pt]
        \hline
            \end{array}
    \end{math}
\caption{Coefficients $a_{CC'}$ characterizing the transition amplitudes for $\mu\to e \gamma$, $\tau\to e \gamma$ and $\tau\to \mu\gamma$, in the MI approximation in which $n_0$ and $n_i^c$ are set to one and $\hat{n}_{1,2}$ to zero and by taking the limit the $\mu=M_{1,2}=m_{SUSY}$.  Numerical values are given in units of $g^2/(192 \pi^2)$ and using $\sin^2\theta_W=0.23$, $(z_i/y_i+\zeta)=1$, and $\tan\beta=2\div 15$.}
\label{table_aCC}
\end{table}
As one can see, in this limit the dominant coefficient is  ${a}_{LL}$, which is larger than $a_{RL}=a_{LR}$ by a factor $7\div 54$, 
and larger than $a_{RR}$ by a factor $-(4\div 13)$, depending on  $\tan\beta=2\div 15$. Assuming coefficients $w_{ij}^{CC'}$ of order one
in eqs. (\ref{wcoeff}), we see that the most important contributions in the amplitudes for the considered processes are $a_{RL} u$ and $a_{LL} u^2$. The ratio between the sub-leading and the leading one is $(a_{LL}/a_{RL}) u\approx (7\div 54) u$.
When $u$ is close to its lower bound, which in our model requires a small value of $\tan\beta$, the leading contribution clearly dominates
over the sub-leading one. However, for $u$ close to $0.05$, which allows to consider larger values of $\tan\beta\approx 15$, the non-leading
contribution can be as large as the leading one and can even dominate over it. The transition between the two regimes occurs towards larger values
of $u$.

The numerical dominance of the coefficient $a_{LL}$ has also another consequence: for vanishing $w^{RL}_{ij}$, $R_{ij}$ is dominated by the contributions of $a_{LL} w^{LL}_{ij}$, whose values
are not universal, but expected to be of the same order of magnitude for all channels. Thus even when $w^{RL}_{ij}=0$,
we predict $R_{\mu e}\approx R_{\tau\mu}\approx R_{\tau e}$. Even if to illustrate our results explicitly we have taken the special limit $\mu=M_{1,2}=m_{SUSY}$, 
a numerical study confirms that they remain approximately correct when more generic regions of the parameter space of the model are considered.
\subsection{Failure of the effective Lagrangian approach}
Why the results of the effective Lagrangian approach in the SUSY case do not apply in the model under consideration?
It is worth to summarize the main assumptions underlying the effective Lagrangian approach in the SUSY case in \cite{A4eff}:
\begin{itemize}
\item[1.] 
The only sources of chirality flip are either fermion masses or sfermion masses of RL-type.
\item[2.]
Both, fermion masses and sfermion masses of RL-type, up to the order $u^2$, are dominated by the insertions of $\varphi_T$ or $\varphi_T^2$ in the relevant operators.
\end{itemize}
As a consequence also the operators describing dipole transitions, which have the same chiral structure as the charged lepton mass terms, 
are dominated by the insertions of  $\varphi_T$ or $\varphi_T^2$, up to the order $u^2$. For instance one such operator is:
\be
\dd\frac{e}{m_{SUSY}^2} \dd\frac{\theta_{FN}^2}{\Lambda_f^2} e^c H_d \sigma_{\mu\nu} F^{\mu\nu} \left(\dd\frac{\varphi_{T}}{\Lambda_f} l\right)+...
\label{OLD}
\ee
and similarly for $\mu^c$ and $\tau^c$. Dots denote additional contributions such as those coming from the insertion of $\varphi_T^2$.
Here $e$ is the electric charge.

Indeed, in our model condition 2 is violated. The explicit SUSY model considered here contains another set of fields which was not present in the effective Lagrangian framework,
namely the driving fields ${\varphi_0^T}$. The driving fields represent an important ingredient of our model, since they are directly responsible for the vacuum alignment
of the flavon fields. In the effective Lagrangian approach the alignment was postulated, without referring to a specific dynamical mechanism to generate it, and
the driving fields were not included among the relevant degrees of freedom. In our model ${\varphi_0^T}$ has no direct coupling to matter fields. 
Such a coupling arises indirectly through the mediation of the flavon $\varphi_T$,
which is coupled to matter via interactions suppressed by $1/\Lambda_f$ and to ${\varphi_0^T}$ via interactions proportional to e.g.
a large scale $M_T$:
\be
w=x_e \dd\frac{\theta_{FN}^2}{\Lambda_f^2} e^c H_d \left(\dd\frac{\varphi_T}{\Lambda_f} l\right) + M_T ({\varphi_0^T} \varphi_T)+...
\ee
$M_T$ is a mass scale of the order of the VEV of $\varphi_T$, $M_T\approx u~ \Lambda_f$ and dots denote further contributions. 
It can be shown that the VEV of ${\varphi_0^T}$ is proportional to $m_{SUSY}$ \cite{VEVsaux}.
When the auxiliary fields of $\varphi_T$ are eliminated slepton masses of RL-type receive a contribution from the insertion of $\ov{\varphi}_0^T$:
\be
y_e \dd\frac{\theta_{FN}^2}{\Lambda_f^2}~ \tilde{e}^c H_d \left(M_{T} \dd\frac{\ov{\varphi}_0^T}{\Lambda_f} \tilde{l}\right)+...
\ee
Similar terms for $\tilde{\mu}^c$ and $\tilde{\tau}^c$ are also present. Therefore dipole operators with the same type of insertions are expected in the low-energy limit. Up to loop factors:
\be
\dd\frac{e}{m_{SUSY}^3} \dd\frac{\theta_{FN}^2}{\Lambda_f^2} e^c H_d \sigma_{\mu\nu} F^{\mu\nu} \left(M_T \dd\frac{\ov{\varphi}_0^T}{\Lambda_f} l\right)+...
\label{NEW}
\ee
Since ${\varphi_0^T}$ has a dominant VEV of order $m_{SUSY}$ and $M_T/\Lambda_f\approx u$, this operator is similar in size to the operator
of eq. (\ref{OLD}). 
A similar contribution arises, if the coupling of two flavons $\varphi_T$ to the driving field $\varphi_0^T$ is taken into account
instead of the term $M_T (\varphi^T_0 \varphi_T)$.
Operators of this type were not included in the effective Lagrangian approach, simply because the driving field $\varphi_0^T$
was absent.
The set of operators of type as in eq. (\ref{NEW}) and eq. (\ref{OLD}) are aligned in flavour space only to the LO. At the NLO such an alignment fails
and this produces a non-vanishing term $w^{RL}_{ij}$. Notice that the insertion of ${\varphi_0^T}$ both in the slepton mass and in the dipole operators
occurs through a combination of the type $M_{T}\ov{\varphi}_0^T/\Lambda_f$. If we did not account for the large coupling of order $M_T$, the mere insertion of $\ov{\varphi}_0^T/\Lambda_f$ in the mass and in the dipole operators would largely underestimate the effect.
It would be interesting to investigate the low-energy structure of dipole operators in a complete SUSY model in which the driving fields are absent
and the alignment of flavon vacua is realized via some alternative mechanism. This would allow to verify whether 
the failure of the effective Lagrangian approach is entirely due to the presence of the driving fields or other subtleties occur when performing the low-energy limit.

\section{Numerical analysis}

In this section we perform a numerical study of the normalized branching ratios $R_{ij}$
and of the deviation $\delta a_\mu$  of the anomalous magnetic moment of the muon from the
SM value. We use the full one-loop results for the branching ratios of the radiative decays
as well as for $\delta a_\mu$. These can be found in \cite{deltaamu_susy,HisanoFukuyama,Arganda,g2BR} and are displayed in appendix C for 
convenience.

\subsection{Framework}

As discussed in the preceding sections, in our model the flavour symmetry $A_4 \times Z_3 \times \Uu_{FN} \times \Uu_R$ 
constrains not only the mass matrices of leptons, but also those of sfermions. These are given at the high energy
scale $\La_f \approx \La_L$, which we assume to be close to $10^{16}$ GeV, the SUSY grand unification scale.
The flavour symmetry does not fix the soft SUSY mass scale $m_{SUSY}$.
It also does not constrain the parameters involved in the gaugino as well as the Higgs(ino) sector. These are fixed by
our choice of a SUGRA framework in which $m_{SUSY}$ is the common soft mass scale for all scalar particles 
(in the literature usually denoted as $m_0$)
and $m_{1/2}$ the common mass scale of the gauginos. Thus, at the scale
$\La_f \approx \La_L$ we have 
\footnote{The gluino mass parameter $M_3$ is not relevant in our analysis.}
\beq
M_1 (\La_L) = M_2 (\La_L) = m_{1/2} \; .
\eeq
Effects of RG running lead at low energies (at the scale $m_W$ of the $W$ mass) to the following masses for gauginos
\beq
M_1(m_W)\simeq\dfrac{\alpha_1(m_W)}{\alpha_1(\Lambda_L)}M_1(\Lambda_L)\qquad\qquad
M_2(m_W)\simeq\dfrac{\alpha_2(m_W)}{\alpha_2(\Lambda_L)}M_2(\Lambda_L)\;,
\eeq
where $\alpha_i=g_i^2/4\pi$ ($i=1,2$) and according to gauge coupling unification at $\La_f\approx \La_L$, $\alpha_1(\Lambda_L)=\alpha_2(\Lambda_L)\simeq 1/25$.
Concerning the effects of the RG running on the soft mass terms, as we have seen in section 3.2 these are small or can be
absorbed into our parametrization of the soft mass terms. 
Thus, in the contributions $(\hat{m}_{eRL}^2)_{1,2}$ to the RL block we take $m_{SUSY}$ as input parameter. 
Nevertheless, we explicitly take into account the RG effect on the average mass scale of the LL block, $m_L^2$, and in the RR block, $m_R^2$,
\beq
\begin{array}{rcl}
m_L^2(m_W)&\simeq& m_L^2(\Lambda_L)+0.5M_2^2(\Lambda_L)+0.04M_1^2(\Lambda_L) \simeq m_{SUSY}^2 +0.54 m_{1/2}^2 \; ,\\[3mm]
m_R^2(m_W)&\simeq& m_R^2(\Lambda_L)+0.15M_1^2(\Lambda_L) \simeq m_{SUSY}^2 +0.15 m_{1/2}^2 \; .
\end{array}
\eeq
The parameter $\mu$ is fixed through the requirement of correct electroweak symmetry breaking
\beq
|\mu|^2=\dfrac{m_{H_d}^2-m_{H_u}^2\tan^2\beta}{\tan^2\beta-1}-\dfrac{1}{2}m_Z^2\;.
\label{defmu}
\eeq
Since in the SUGRA framework the soft Higgs mass parameters are also given by $m_{SUSY}$ at the high energy scale,
 $m_{H_u}^2 (\La_L)=m_{H_d}^2 (\La_L)=m_{SUSY}^2$ , the relation in eq. (\ref{defmu}) reads at low energies
\beq
|\mu|^2\simeq-\dfrac{m_Z^2}{2}+m_{SUSY}^2\dfrac{1+0.5\tan^2\beta}{\tan^2\beta-1}+m_{1/2}^2\dfrac{0.5+3.5 \tan^2\beta}{\tan^2\beta-1}\;,
\label{defmuSUGRA}
\eeq
so that $\mu$ is determined by $m_{SUSY}$, $m_{1/2}$  and $\tan\beta$ up to its sign. 
We recall that in our model the low energy parameter $\tan\beta$ is related to the size of the expansion parameter $u$, the mass of
the $\tau$ lepton and the $\tau$ Yukawa coupling $y_\tau$, as shown in eq. (\ref{tanb&u&yt}). For this reason, requiring $1/3 \lesssim |y_\tau| \lesssim 3$ 
constrains $\tan\beta$ to lie in the range $2 \lesssim \tan\beta \lesssim 15$. As already commented, the lower bound $\tan\beta =2$
is almost excluded experimentally, since such low values of $\tan\beta$ usually lead to a mass for the lightest Higgs below the LEP2 bound
of $114.4$ GeV \cite{mhbound}. \footnote{This bound assumes that the Higgs is SM-like. For the case of generic MSSM Higgs the bound is much lower, $91.0$ GeV
\cite{MSSMmhbound}.}

In our numerical analysis the parameters are the following: the two independent mass scales $m_{SUSY}$ and $m_{1/2}$,
the sign of the parameter $\mu$ and the parameters of the slepton mass matrices shown in section 3.1 in the physical basis.
We recall that the results of section 3.1 have been obtained under the assumption that the parameters are real
and we keep working under the same assumption here.  We also assume that the parameters on the diagonal of the slepton mass matrices $(\hat{m}_{(e,\nu)LL}^2)_K$ and $(\hat{m}_{eRR}^2)_K$,
$n_0$ and $n^c_{1,2,3}$,
are positive in order to favour positive definite square-masses, to avoid electric-charge breaking minima and 
further sources of electroweak symmetry breaking.
The absolute value of the ${\cal O}(1)$ parameters is varied between $1/2$ and $2$. 
We will chose some representative values for $u$ in the allowed range $0.007 \lesssim u \lesssim 0.05$. The other expansion parameter $t$ is fixed to be $0.05$. 
In the analysis of the normalized branching ratios $R_{ij}$ we fix $\tan\beta$ and $u$ and then we derive the Yukawa couplings $y_e$, $y_\mu$ and $y_\tau$.
When discussing the anomalous magnetic moment of the muon instead we vary $y_\tau$ between $1/3$ and $3$ and 
calculate $\tan\beta$ by using eq. (\ref{tanb&u&yt}). Having determined $\tan\beta$, the Yukawa couplings $y_e$ and $y_\mu$ can be computed.

The allowed region of the parameter space is determined by performing several tests. We check
whether the mass of the lightest chargino is above 100 GeV \cite{pdg2008}, whether
the lightest neutralino is lighter than the lightest charged slepton, whether the lower bounds for the charged slepton masses are obeyed \cite{pdg2008} 
and whether the masses of all sleptons are positive. The constraint on the mass of the lightest chargino implies
a lower bound on $m_{1/2}$ which slightly depends on the sign of $\mu$. In our plots for $R_{ij}$
we also show the results for points of the parameter space that do not respect the chargino mass bound.
For low values of $m_{SUSY}$, e.g. $m_{SUSY}=100$ GeV, the requirement that the lightest neutralino is lighter than the
lightest charged slepton is equivalent to the requirement that the parameters in the
diagonal entries of the slepton mass matrices $(\hat{m}_{(e,\nu)LL}^2)_K$ and $(\hat{m}_{eRR}^2)_K$ are larger than one. 
For larger values of $m_{SUSY}$, e.g. $m_{SUSY}=1000$ GeV, this requirement does not affect our analysis anymore.
We note that masses of charginos and neutralinos  are essentially independent from the ${\cal O}(1)$ parameters of the slepton mass matrices and thus their masses fulfill with very good accuracy (better for larger $m_{1/2}$)
\beq
M_{ \widetilde{\chi}^0_1} \approx 0.4 m_{1/2} \; , M_{ \widetilde{\chi}^0_2} \approx M_{ \widetilde{\chi}^-_1} \approx 0.8 m_{1/2} \; ,
M_{ \widetilde{\chi}^0_3} \approx M_{ \widetilde{\chi}^0_4} \approx  M_{ \widetilde{\chi}^-_2} \approx |\mu| \; .
\eeq
For the slepton masses we find certain ranges which depend on our choice of the ${\cal O}(1)$ parameters.

\subsection{Results for radiative leptonic decays}

\begin{figure}[h!]
 \centering
\subfigure[$\tan\beta=2$, $u=0.01$ and $m_{SUSY}=100$ GeV.]
   {\includegraphics[width=8cm]{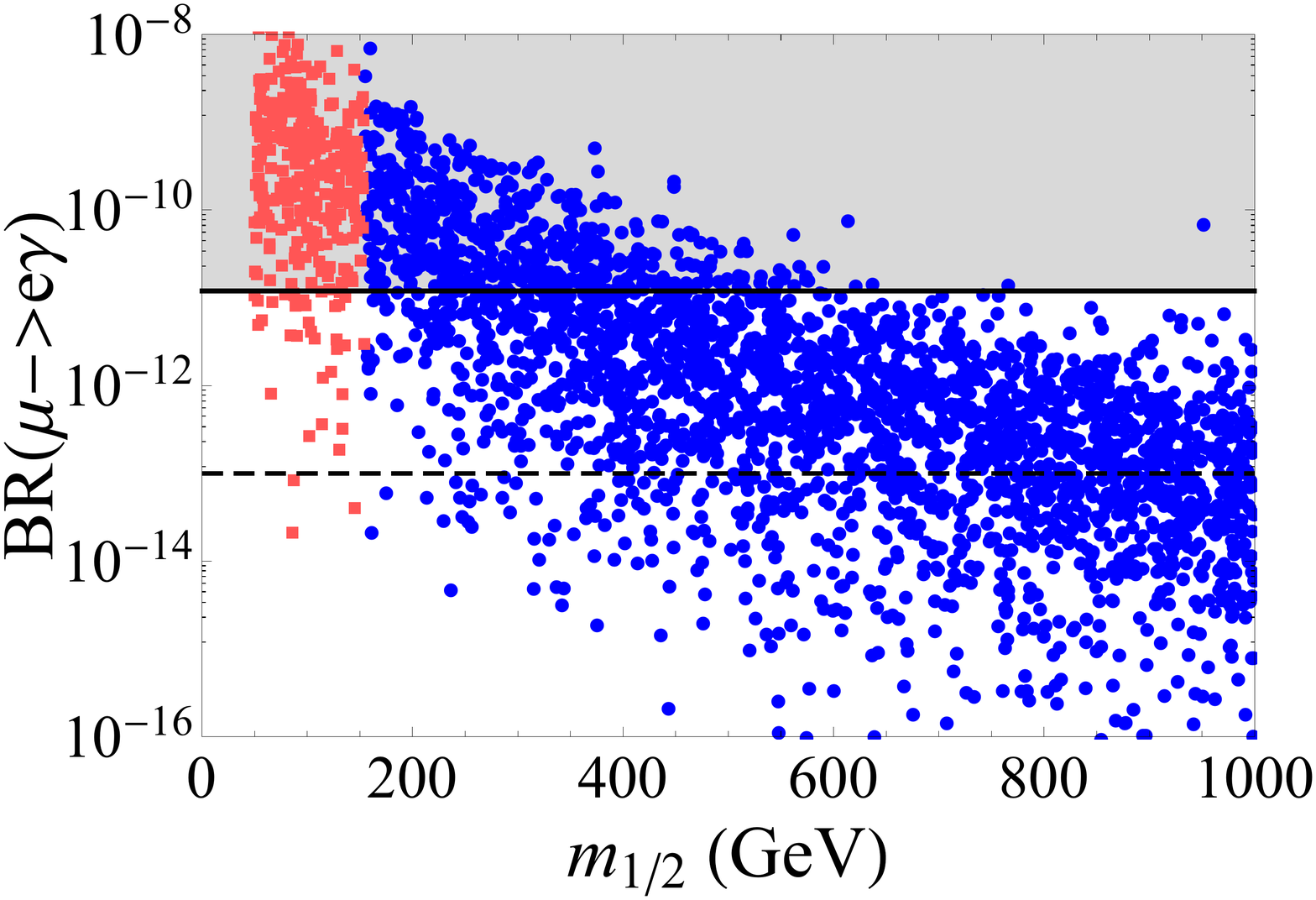}}
\subfigure[$\tan\beta=2$, $u=0.01$ and $m_{SUSY}=1000$ GeV.]
   {\includegraphics[width=8cm]{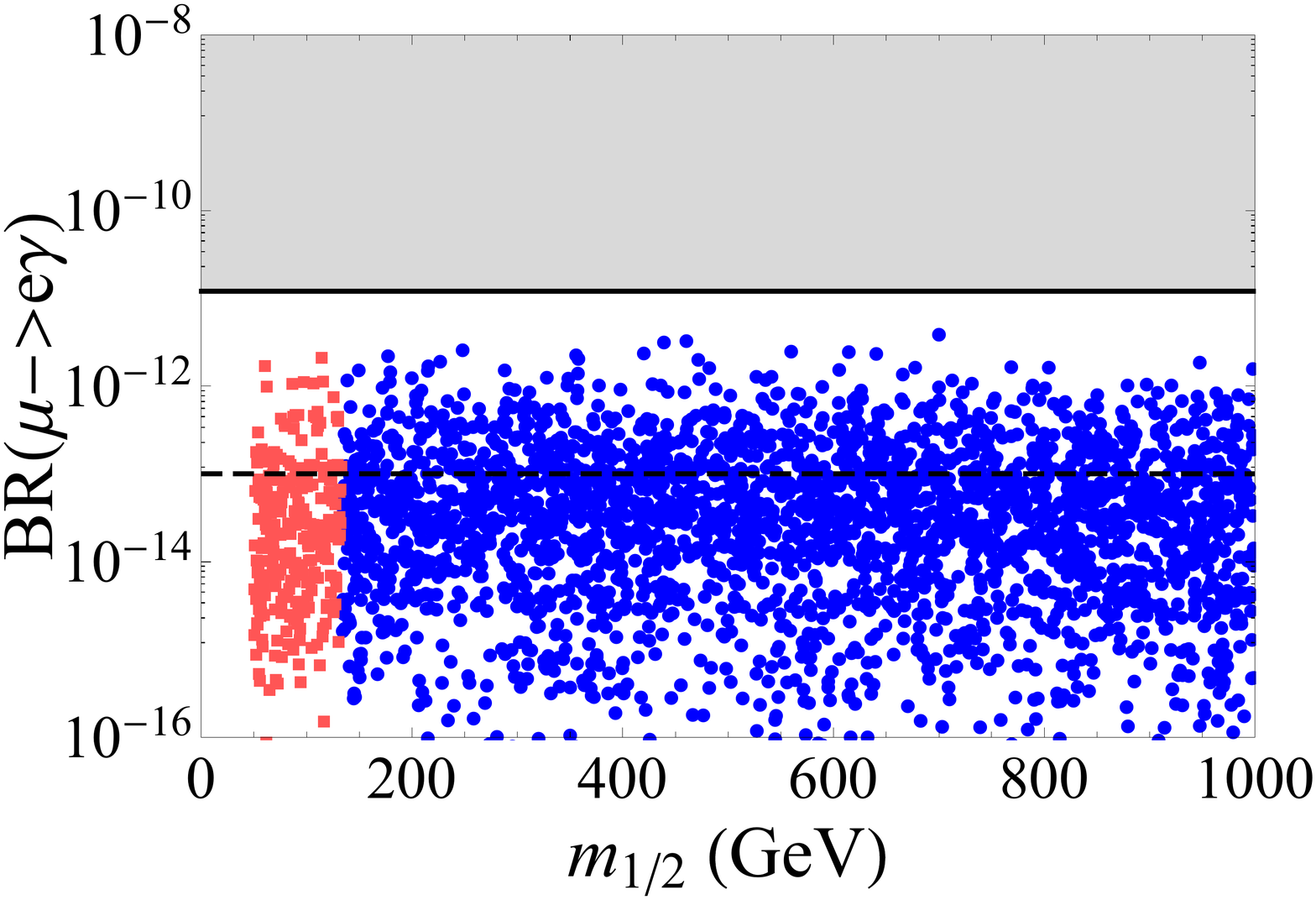}}
\subfigure[$\tan\beta=2$, $u=0.05$ and $m_{SUSY}=100$ GeV.]
   {\includegraphics[width=8cm]{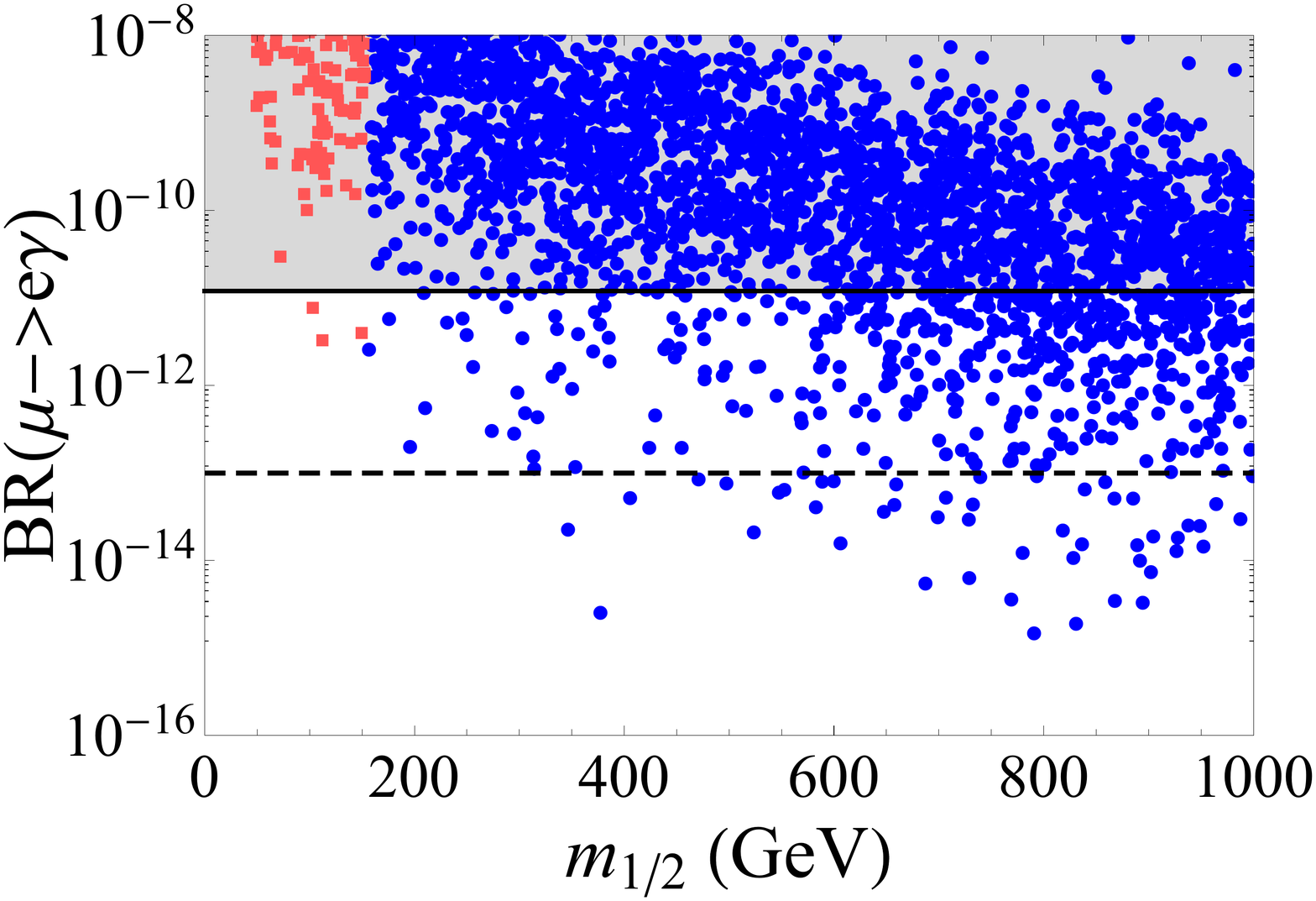}}
\subfigure[$\tan\beta=2$, $u=0.05$ and $m_{SUSY}=1000$ GeV.]
   {\includegraphics[width=8cm]{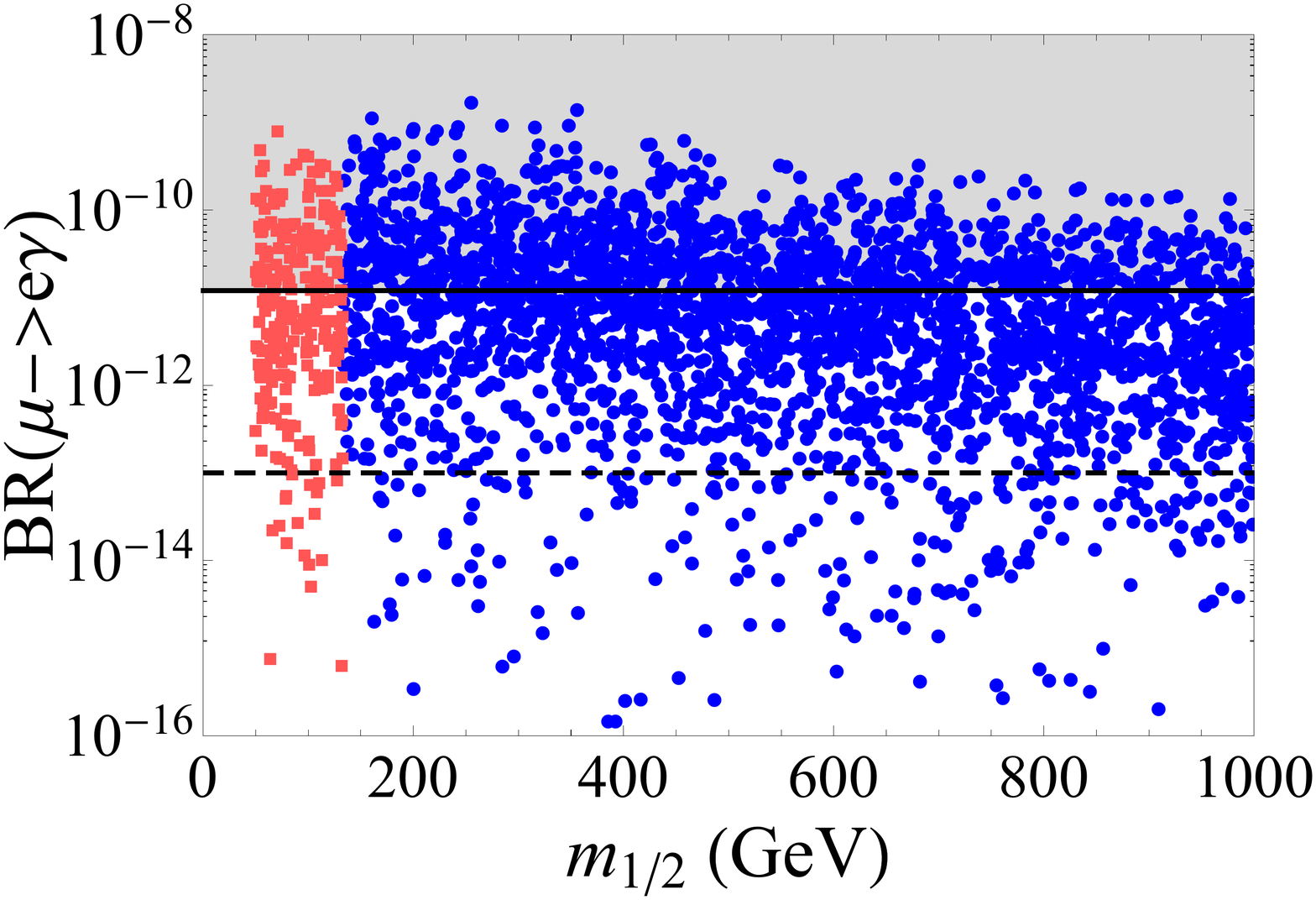}}
\subfigure[$\tan\beta=15$, $u=0.05$ and $m_{SUSY}=100$ GeV.]
   {\includegraphics[width=8cm]{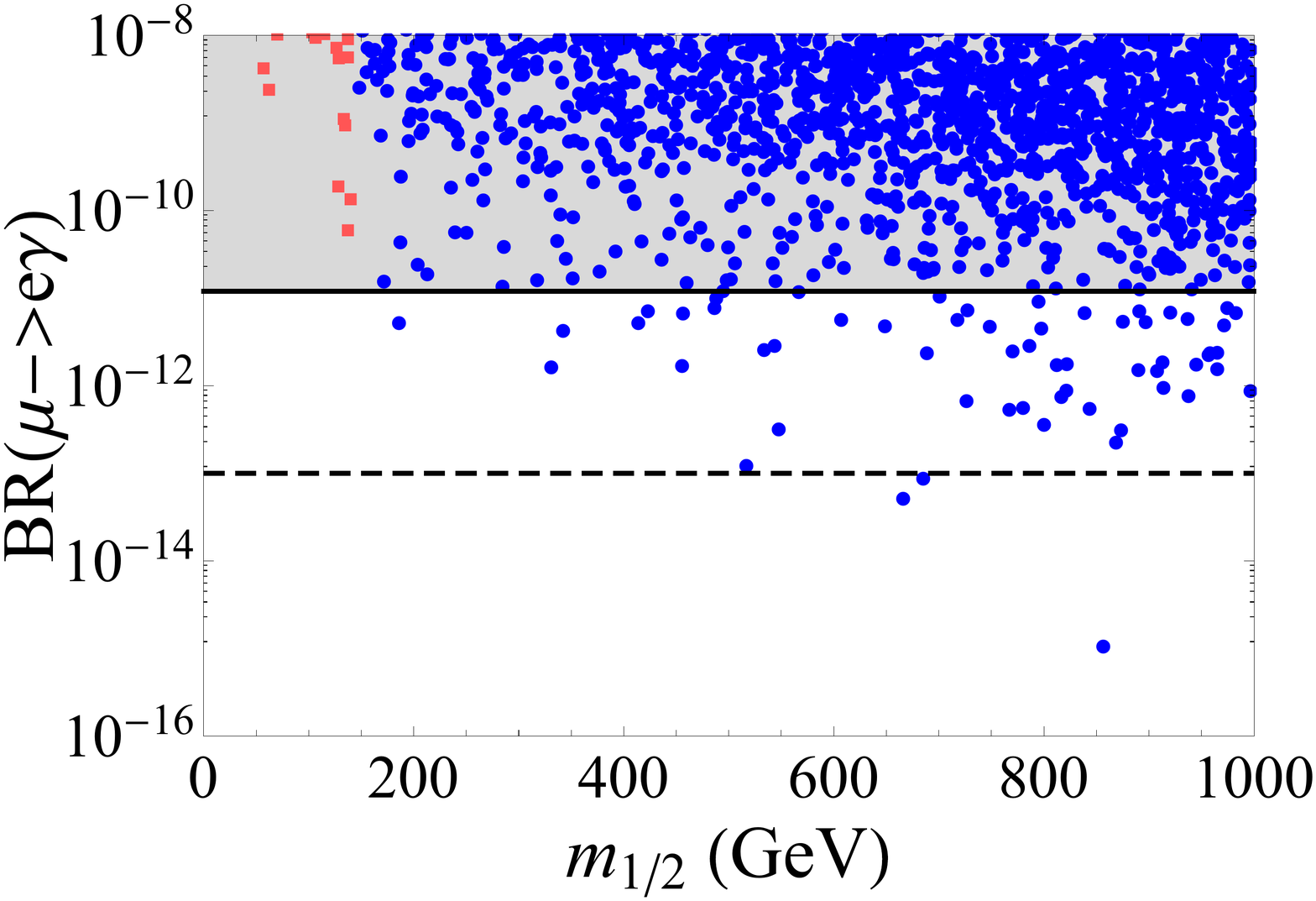}}
\subfigure[$\tan\beta=15$, $u=0.05$ and $m_{SUSY}=1000$ GeV.]
   {\includegraphics[width=8cm]{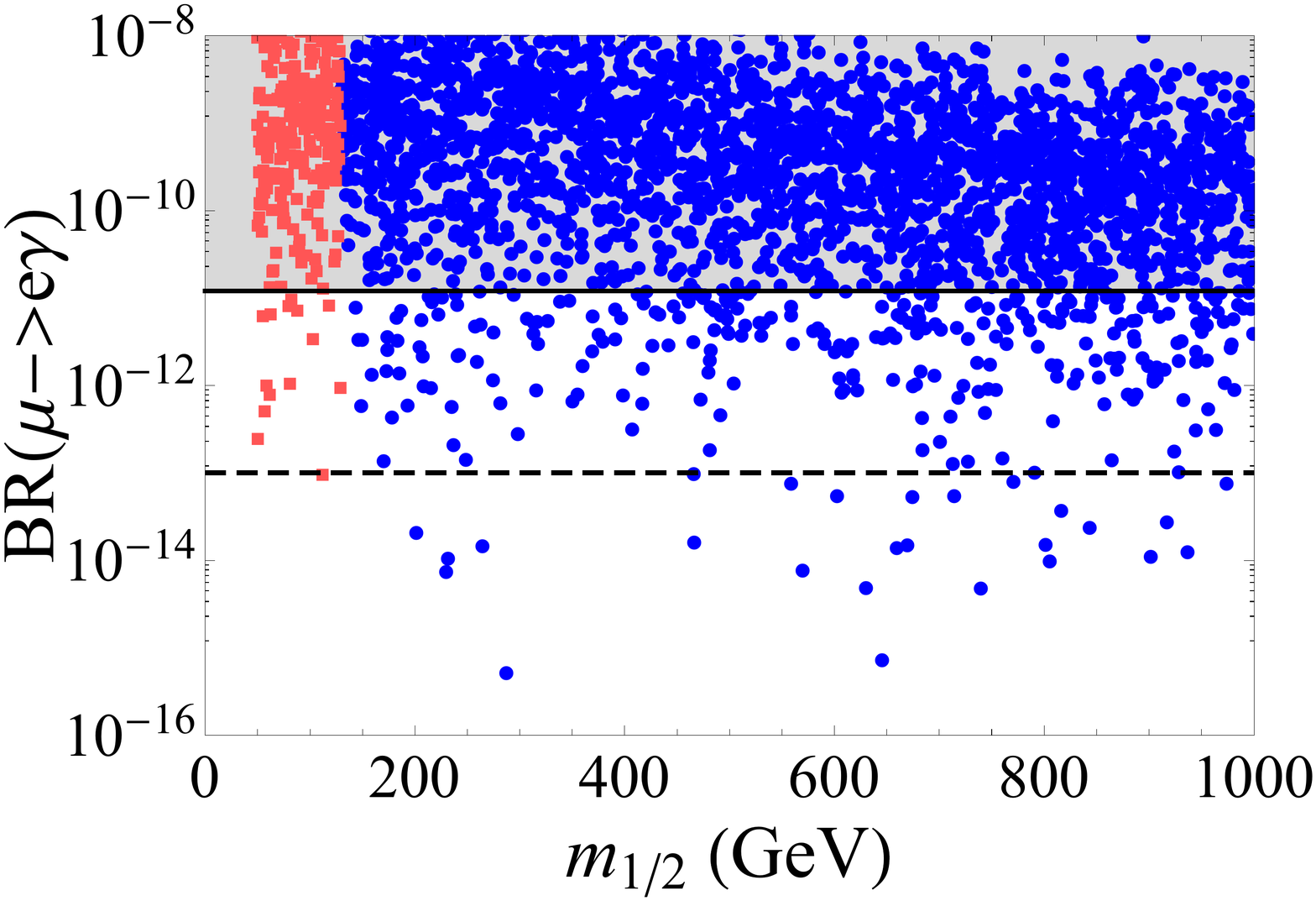}}
   \caption{Scatter plots of $BR(\mu\to e \gamma)$ as a function of $m_{1/2}$, for different values of $\tan\beta$, $u$ and $m_{SUSY}$. 
The red (dark gray) points correspond to points in which the mass of the lightest chargino is below the limit coming from direct searches.
The horizontal lines show the current MEGA bound (continuous line) and the prospective MEG bound (dashed line).}
\label{ScatterBR}
\end{figure}

We first discuss the results for the branching ratio of the decay $\mu\to e\gamma$. This branching ratio is severely constrained
by the result of the MEGA experiment \cite{MEGA}
\beq
R_{\mu e} \approx BR(\mu\to e\gamma) < 1.2 \times 10^{-11}
\eeq
and will be even more constrained by the on-going MEG experiment \cite{MEG} which will probe the regime
\beq
R_{\mu e}\approx BR(\mu\to e\gamma) \gtrsim 10^{-13} \; .
\eeq
We explore the
parameter space of the model by considering two different values of the expansion parameter $u$, $u=0.01$ and $u=0.05$, two different
values of $\tan\beta$, $\tan\beta=2$ and $\tan\beta=15$, as well as two different values of the mass scale $m_{SUSY}$, $m_{SUSY}=100$ GeV
and $m_{SUSY}=1000$ GeV. We show our results in scatter plots in Figure \ref{ScatterBR} choosing $m_{1/2}$ to be $m_{1/2} \lesssim 1000$ GeV. 
All plots shown in Figure \ref{ScatterBR} are generated for $\mu>0$. 

As one can see from Figure \ref{ScatterBR}(a), 
for very low $\tan\beta=2$, small $u=0.01$, small $m_{SUSY}=100$ GeV the experimental upper limit from the MEGA experiment 
on $BR(\mu\to e\gamma)$ can be passed in almost all parameter space of our model
for values of $m_{1/2}$ as small as $450$ GeV. For $m_{SUSY}=100$ GeV and $m_{1/2}=450$ GeV
the sparticle masses are rather light: the lightest neutralino has a mass of $175$ GeV, the lightest chargino of $350$ GeV, the masses
of the right-handed (charged) sleptons vary between $175$ and $285$ GeV and the masses of the left-handed sleptons are in the range $(250\div 500)$
GeV. Thus, especially the right-handed sleptons are expected to be detected at LHC. In a model also including quarks (and hence squarks)
we find for the squarks that they can have masses $\gtrsim 700$ GeV and gluinos with masses of about $1000$ GeV, all accessible at LHC.
To pass the prospective bound coming from the MEG experiment in a sizable portion of parameter space of our model
$m_{1/2}$ has to be chosen larger, $m_{1/2} \gtrsim 600$ GeV. Then, however,
the masses of the sleptons might only be detected at LHC in case of right-handed sleptons. 
As one can see, values of $m_{1/2} \lesssim 155$ GeV are excluded due to
the constraint on the lightest chargino mass. Studying the same value of $\tan\beta$ and $u$, but taking $m_{SUSY}$ to be as large as 
$1000$ GeV, we can see from Figure \ref{ScatterBR}(b) that now the bound set by the MEGA experiment on $BR(\mu\to e\gamma)$ is respected 
in the whole parameter space of our model
for all values of $m_{1/2}$. Also the foreseen limit of the MEG experiment can only exclude a smaller part of the parameter space of
the model for all values of $m_{1/2}$. 
In this setup, the prospects for detecting SUSY particles at LHC are the best for gauginos due to the possible low value of $m_{1/2}$.
The slepton masses are expected to be roughly $m_{SUSY}$ and thus too large to allow for a detection at LHC.

Increasing the
value of the expansion parameter $u$ from $u=0.01$ to $u=0.05$, as done in Figure \ref{ScatterBR}(c) and \ref{ScatterBR}(d), increases
also the branching ratio of the decay $\mu\to e\gamma$ by approximately two orders of magnitude, since the different contributions to the branching ratio 
scale at least with $u^2$, as analyzed in section 4. For this reason for low values of $m_{SUSY}=100$ GeV, $m_{1/2}$ has to take values
$m_{1/2} \gtrsim 600$ GeV in order for the result of $BR(\mu\to e\gamma)$ to be compatible with the MEGA bound at least in some portion
of the parameter space of our model. For the point $(m_{SUSY},m_{1/2})=(100 \, \rm GeV,
600 \, GeV)$ the sparticle spectrum is characterized as follows: the lightest neutralino has a mass of $240$ GeV, the lightest chargino of
$470$ GeV, right-handed sleptons between $250$ and $350$ GeV and left-handed sleptons generally above $300$ GeV. For this reason
there still exists the possibility to detect right-handed sleptons at LHC. Concerning gluinos and squarks these are expected to
have masses between $1000$ and $1500$ GeV so that they also can be detected at LHC. As one can see from Figure \ref{ScatterBR}(c) the on-going
MEG experiment can probe nearly the whole parameter space of the model for $\tan\beta=2$, $u=0.05$ and $m_{SUSY}=100$ GeV for values
of $m_{1/2} \lesssim 1000$ GeV. Increasing the parameter $m_{SUSY}$ to $1000$ GeV shows that applying the existing bound on
$BR(\mu\to e\gamma)$ of $1.2 \times 10^{-11}$ cannot exclude small values of $m_{1/2}$. The situation changes, if the expected bound 
from the MEG experiment is employed, because then values of $m_{1/2}$ smaller than $1000$ GeV become disfavoured. 

Finally, we show in Figure \ref{ScatterBR}(e) and \ref{ScatterBR}(f) the results obtained for $\tan\beta=15$. We remind that this value is
the largest possible one of $\tan\beta$ in our model. Requiring that the $\tau$ Yukawa coupling does not become too large entails that
$\tan\beta=15$ fixes the expansion parameter $u$ to take a value close to its upper limit, $u=0.05$. The value of $BR(\mu\to e \gamma)$ 
is thus enhanced through $\tan\beta$ as well as $u$. This is clearly shown in  Figure \ref{ScatterBR}(e) and \ref{ScatterBR}(f), because
for a low value of $m_{SUSY}=100$ GeV already the MEGA bound practically excludes almost the whole parameter space of our model
for all values of $m_{1/2} \lesssim 1000$ GeV.
 Increasing the mass parameter $m_{SUSY}$ to $1000$ GeV slightly improves the situation, because
now there exists a marginal probability to pass the MEGA bound. Again, however, the MEG experiment 
can probe all parameter space of our model for $m_{1/2} \lesssim 1000$ GeV. Thus, for $m_{SUSY} \lesssim 1000$ GeV and $m_{1/2} \lesssim 1000$ GeV
 the parameter space of our model is already severely constrained for moderate values of $\tan\beta$ which entail large $u \approx 0.05$ 
by the bound coming from the MEGA collaboration, but surely will be conclusively probed by the MEG experiment.
Choosing $\mu<0$ hardly affects the results presented here apart from slightly decreasing the lower bound on $m_{1/2}$ coming from the
chargino mass bound. 
Thus, all statements made also apply for $\mu<0$.

In summary, the current bound on $BR(\mu\to e\gamma)$ prefers regions in the parameter space of our model with small $u$ or
small $\tan\beta$, as long as the SUGRA mass parameters should be chosen smaller than $1000$ GeV. The foreseen MEG bound
strongly favours regions in which $u$ is small for $m_{SUSY}$ and $m_{1/2}$ being not too large. The fact that smaller 
values of $u$ are preferred has consequences also for the expectations of the detection prospects for the reactor mixing angle $\theta_{13}$,
because this angle scales with $u$: it might thus not be possible to detect $\theta_{13}$ with the reactor and neutrino beam experiments 
under preparation \cite{Schwetz:2007my,Theta13future}.

\vspace{0.1in}

Concerning the radiative $\tau$ decays, $\tau\to \mu\gamma$ and $\tau\to e\gamma$, the result found in the MI approximation that 
the branching ratios of these decays are of the same order of magnitude as $BR(\mu\to e\gamma)$ is essentially confirmed in a numerical
analysis. Due to the random parameters differences up to two orders of magnitude are expected and found, especially for the case of larger
$\tan\beta$. However, it is still
highly improbable that the decays $\tau\to \mu\gamma$ and $\tau\to e\gamma$ could be detected at a SuperB factory, assuming a prospective
limit of $BR(\tau\to \mu \gamma), BR(\tau\to e \gamma) \gtrsim 10^{-9}$ \cite{SuperB}.

\subsection{Results for anomalous magnetic moment of the muon}

As is well known, the value found for the anomalous magnetic moment of the muon \cite{amuexp}
\be
a_\mu^{EXP}=116592080(63)\times 10^{-11}
\ee
shows a 3.4 $\sigma$ deviation 
\be
\delta a_\mu=a_\mu^{EXP}-a_\mu^{SM}=+302(88)\times 10^{-11}
\label{Da}
\ee
from the value expected in the SM \cite{amuSM} 
\be
a_\mu^{SM}=116591778(61)\times 10^{-11} \; .
\ee
Similar discrepancies have been reported in \cite{hagiwara}. Thus, it might be interesting to consider
the case in which this deviation is attributed to the presence of SUSY particles with masses of a few hundred GeV.
The one-loop contribution to the anomalous magnetic moment of the muon in supersymmetric extensions of the SM has been studied by several authors \cite{deltaamu_susy}.
\begin{figure}[h!]
 \centering
\subfigure[$u=0.01$.]
   {\includegraphics[width=8cm]{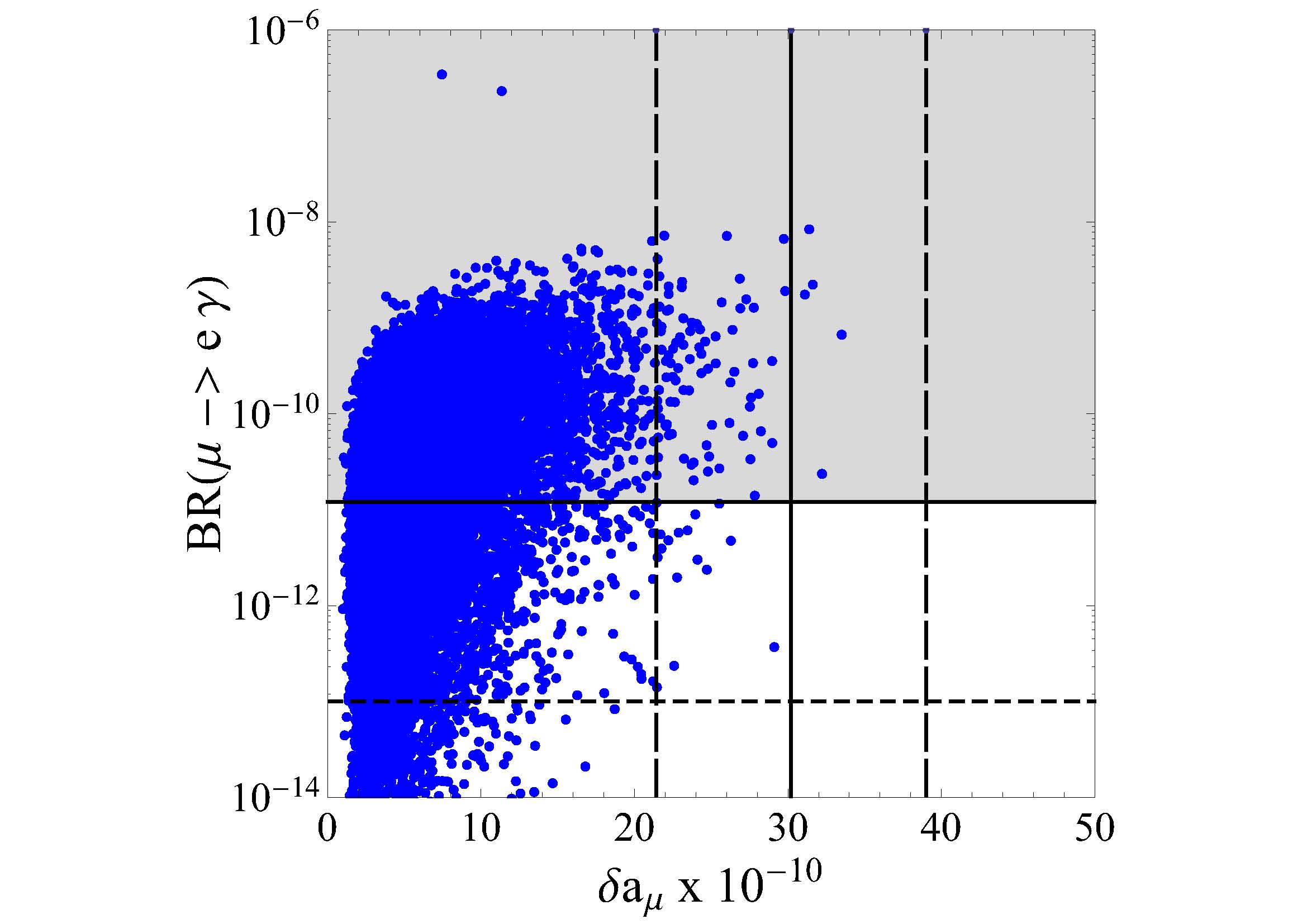}}
\subfigure[$u=0.05$.]
   {\includegraphics[width=8cm]{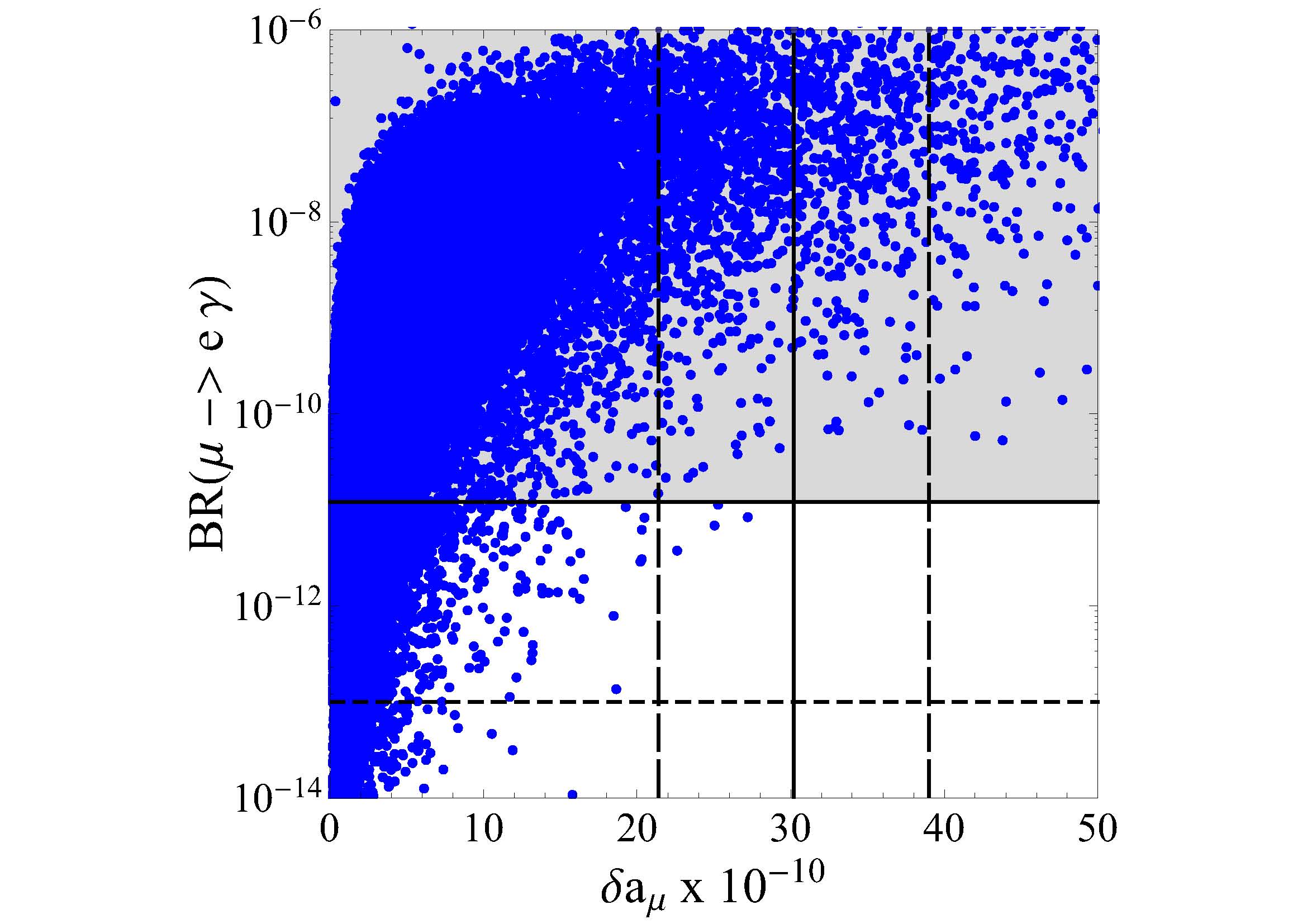}}
   \caption{Scatter plots in the plane $BR(\mu\to e\gamma)-\delta a_\mu$, for values of $u=0.01,\,0.05$. The value of $\tan\beta$ is fixed through 
the relation with the $\tau$ Yukawa coupling, which lies in the interval $[1/3,3]$. The values of $m_{SUSY}$ and of $m_{1/2}$ are chosen between 
$10$ and $300$ GeV for the left panel and between $10$ and $1000$ GeV in the right one. The horizontal lines correspond to the MEGA 
(continuous line) and the MEG bounds (dashed line); 
the vertical lines correspond to the measurements on $\delta a_\mu$: the continuous one is the best fit value and 
the dashed ones correspond to the $3\sigma$ boundaries.}
\label{ScatterBrG2}
\end{figure}
We study the compatibility between the requirement that $\delta a_\mu$ is explained by the exchange of relatively light SUSY particles
and the experimental upper limit on $BR(\mu\to e\gamma)$ coming from the MEGA
experiment. We choose again two different values of $u$, $u=0.01$ and $u=0.05$. To better explore
the parameter space we vary the $\tau$ Yukawa coupling between $1/3$ and $3$
and fix the value of $\tan\beta$ through the relation given in eq. (\ref{tanb&u&yt}).
As a consequence in the plot for $u=0.01$, see Figure \ref{ScatterBrG2}(a),
$2 \lesssim \tan\beta \lesssim 3$ holds and for $u=0.05$ $\tan\beta$ takes values $2\lesssim \tan\beta \lesssim 15$. Similarly, $m_{SUSY}$
and $m_{1/2}$ are chosen to lie in intervals $[10 \, \rm GeV, 300 \, GeV]$ and $[10 \, \rm GeV, 1000 \, GeV]$ for $u=0.01$ and $u=0.05$, respectively. 
The different choice of intervals is due to the fact that values of a few hundred GeV for $m_{SUSY}$ and $m_{1/2}$ 
are disfavoured by the existing limit on $BR(\mu\to e\gamma)$ when $u=0.05$. As one can clearly see from Figure \ref{ScatterBrG2}, in almost
the whole parameter space of our model it is not natural to reproduce the observed deviation of the muon anomalous
magnetic moment and at the same time to respect the existing bound on the branching ratio of $\mu\to e \gamma$.

This kind of incompatibility is well known
in supersymmetric theories, because the explanation of the $3.4 \sigma$ discrepancy necessitates small values of $m_{SUSY}$ and $m_{1/2}$ and larger
values of $\tan\beta$, which
in turn enhance the branching ratio of the radiative LFV decays. Thus, we have to conclude that either there exist further sources of contributions
to the anomalous magnetic moment of the muon beyond those present in our model, or - as is also discussed in the literature - the theoretical
 value $a_\mu^{SM}$ found in the SM is closer to $a_\mu^{EXP}$ so that the eventual discrepancy 
 becomes less than $100 \times 10^{-11}$, a value which could well be explained in our model.


\section{Conclusion}

While awaiting the start of LHC and the first exploration of the TeV scale, the solution to the hierarchy problem offered by SUSY
still represents a very appealing option, with several interesting consequences, such as a successful gauge coupling unification,
a viable particle candidate for dark matter and a possible explanation of the observed discrepancy in the anomalous magnetic
moment of the muon. Nevertheless the existence of new states carrying flavour indices at the TeV scale is a serious problem
in model building, given the success of the SM in describing all known phenomena involving flavour transitions. The lepton sector makes no exception.
The flavour conversion observed in neutrino oscillations might be related to potentially large effects in rare decays of the charged leptons.
A naive dimensional estimate of these effects, assuming new physics at the TeV scale, requires a large suppression in the relevant amplitudes, to pass the existing bounds.
A possible explanation for such a suppression is the presence of a flavour symmetry, which is independently motivated by
the hierarchy in the charged lepton mass spectrum and by the nearly TB mixing pattern of the lepton mixing matrix.
Here we have considered a SUSY model invariant under the flavour symmetry group $A_4\times Z_3\times U(1)_{FN}$,
originally proposed in order to describe lepton masses and mixing angles. We have extended the original model
by including SUSY breaking terms consistent with all symmetry requirements.
Our model is an effective theory, valid at energy scales below a cutoff $\Lambda_f \approx \Lambda_L$, where we have derived the spectrum
of SUSY particles, in the slepton sector, under the assumption that the SUSY breaking scale is larger than $\Lambda_f$.
It provides an example of a model in which the slepton mass matrices at the
scale $\Lambda_f$ are not universal. Left-handed sleptons are
approximately universal, with a small departure from universality controlled by $u$, the flavour symmetry breaking parameter.
Right-handed sleptons have soft masses of the same order, but the relative difference among them is expected to be
of order one. Off-diagonal elements in both sectors, as well as in the RL block, are small and have a characteristic pattern
in terms of powers of $u$.
This structure is maintained by the effects coming from the RG running from $\Lambda_f \approx \Lambda_L$ down to the electroweak scale.
The symmetry breaking parameter $u$ lies in a restricted range around few per cent and has a size similar to the reactor mixing angle $\theta_{13}$.
We have exploited the knowledge of the slepton mass matrices to compute the normalized branching ratios $R_{ij}$
for the transitions $\mu\to e \gamma$, $\tau\to\mu\gamma$ and $\tau\to e \gamma$. At variance with other models based on flavour symmetries we found
$R_{\mu e}\approx R_{\tau\mu}\approx R_{\tau e}$ and, given the present limit on $R_{\mu e}$, these rare $\tau$ decays are practically unobservable in our model.
On a more theoretical side, the scaling $R_{ij}\propto u^2$, found in the MI approximation, violates an 
expectation based on an effective Lagrangian approach,
which suggested $R_{ij}\propto u^4$ in the limit of massless final charged lepton. We have identified the source of such a violation
in a single, flavour independent, contribution of the RL block of the slepton mass matrix.
Such a contribution originates from the VEVs of the auxiliary components of the flavon supermultiplets. We have classified the conditions under which
this universal contribution is absent.

In a numerical analysis of $R_{\mu e}$ we found that already the current bound from the MEGA experiment requires the parameter
$u$ to be small or $\tan\beta$ to be small for SUSY mass parameters $m_{SUSY}$ and $m_{1/2}$ below $1000$ GeV, to
guarantee detection of sparticles at LHC. Applying the prospective MEG bound tightens the parameter space of our model even more to 
small $u$ and $\tan\beta$ or requires mass parameters $m_{SUSY}$ and $m_{1/2}$ above $1000$ GeV. Furthermore, we showed that the
deviation of the experimentally observed value of the magnetic moment of the muon from the SM one cannot be naturally explained in our framework, for
$BR(\mu\to e\gamma)$ below the current bound. The maximal value of $\delta a_{\mu}$ in our model is around $100 \times 10^{-11}$ for
$BR(\mu\to e\gamma) \lesssim 10^{-11}$.

\section*{Acknowledgments}
We thank Lorenzo Calibbi, Stefano Rigolin and Fabio Zwirner for many useful discussions.
We recognize that this work has been partly supported by the European Commission under contracts PITN-GA-2009-237920 (UNILHC) and MRTN-CT-2006-035505.
\vfill

\vfill
\newpage


\section*{Appendix A~~~The group $A_4$}
The group $A_4$ is generated by two elements $S$ and $T$ obeying the relations\cite{A4Presentation}:
\be
S^2=(ST)^3=T^3=1~~~.
\ee
It has three independent one-dimensional representations, $1$, $1'$ and $1''$ and one three-dimensional representation $3$.
The one-dimensional representations are given by:
\be
\begin{array}{lll}
1&S=1&T=1~~~,\\
1'&S=1&T=e^{\dd i 4 \pi/3}\equiv\omega^2~~~,\\
1''&S=1&T=e^{\dd i 2\pi/3}\equiv\omega~~~.\\
\label{s$A_4$}
\end{array}
\ee
The three-dimensional representation, in a basis
where the generator $T$ is diagonal, is given by:
\be
T=\left(
\begin{array}{ccc}
1&0&0\\
0&\omega^2&0\\
0&0&\omega
\end{array}
\right),~~~~~~~~~~~~~~~~
S=\frac{1}{3}
\left(
\begin{array}{ccc}
-1&2&2\cr
2&-1&2\cr
2&2&-1
\end{array}
\right)~~~.
\label{ST}
\ee
The multiplication rule for triplet representations is the following:
\be
3\times 3=1+1'+1''+3_S+3_A
\ee
If we denote by:
\be
a=(a_1,a_2,a_3)~~~,~~~~~~~~~~b=(b_1,b_2,b_3)~~~
\ee
two triplets, the singlets contained in their product are given by:
\be
\begin{array}{llll}
1&\equiv(ab)&=&(a_1 b_1+a_2 b_3+a_3 b_2)\\
1'&\equiv(ab)'&=&(a_3 b_3+a_1 b_2+a_2 b_1)\\
1''&\equiv(ab)''&=&(a_2 b_2+a_1 b_3+a_3 b_1)
\label{dec2}
\end{array}
\ee
The two triplets can be separated into a symmetric and an antisymmetric part:
\bea
3_S\equiv(ab)_S&=&\frac{1}{3}(2 a_1 b_1-a_2 b_3-a_3 b_2,2 a_3 b_3-a_1 b_2-a_2 b_1,2 a_2 b_2-a_1 b_3-a_3 b_1)\cr
3_A\equiv(ab)_A&=&\frac{1}{2}(a_2 b_3-a_3 b_2,a_1 b_2-a_2 b_1,a_3 b_1-a_1 b_3)
\label{dec3}
\eea
Moreover, if $c$, $c'$ and $c''$ are singlets transforming as $1$, $1'$ and $1''$, and
$a=(a_1,a_2,a_3)$ is a triplet, then the products $ac$, $ac'$ and $ac''$ are triplets
explicitly given by $(a_1 c,a_2 c, a_3 c)$, $(a_3 c',a_1 c', a_2 c')$ and $(a_2 c'',a_3 c'', a_1 c'')$,
respectively.
Note that due to the choice of complex representation matrices for the
real representation 3 the conjugate $\bar{a}$ of $a \sim 3$ does not
transform as $3$, but rather $(\bar{a}_1, \bar{a}_3, \bar{a}_2)$
transforms as triplet under $A_4$.
The reason for this is that
$T^{\star}= U^{T} T U$ and
$S^{\star}=U^{T} S U =S$ where $U$ is the matrix which exchanges the second and third row and
column.


\mathversion{bold}
\section*{Appendix B~~~Canonical normalization of the kinetic terms and diagonalization of the charged lepton
mass matrix $m_l$}
\mathversion{normal}

We perform the following transformations on the superfields: the kinetic terms are canonically
normalized, i.e. we find the basis in which the matrices $K$ and $K^{c}$, shown in eqs.
(\ref{T},\ref{Tc}), are unit matrices, and in a second step the charged lepton mass matrix $m_l$,
eq. (\ref{yl_subleading}), is diagonalized.
As explained in the main text, we perform these transformations not only on the lepton fields, but also on the
sleptons in order to ensure that the gaugino-lepton-slepton vertices do not violate flavour at this stage.

To diagonalize the
hermitian matrices $K$ and $K^{c}$ we apply the unitary transformations $W$ and $W^{c}$:
\begin{equation}
W^{\dagger} K W = {\tt diag} ~~~~~~~ \mbox{and} ~~~~~~~ (W^c)^{\dagger} K^c W^c = {\tt diag}~~.
\end{equation}
Normalizing $K$ and $K^c$ requires a rescaling of the fields via the real (diagonal)
matrices $R$ and $R^c$:
\begin{equation}
R W^{\dagger} K W R = \mathbb{1} ~~~~~~~~~\mbox{and} ~~~~~~~~~R^c (W^c)^{\dagger} K^c W^c R^c = \mathbb{1}~~.
\end{equation}
The superfields $l$ and $l^c$ ($l^c=(e^c,\mu^c,\tau^c)$ as above) are expressed as:
\begin{equation}
l=W R \, l^{\prime} ~~~~~~~~~\mbox{and} ~~~~~~~~~ l^c = W^c R^c l^{c \, \prime}
\end{equation}
so that the kinetic terms for leptons and sleptons are in their the canonical form,
i.e. eq. (\ref{kt}) becomes:
\[
i \, \bar{l}^{\prime}_i \bar{\sigma}^{\mu} D_\mu l^{\prime}_i
+ i \, \bar{l^c}^\prime_i \bar{\sigma}^{\mu} D_\mu l^{c \, \prime}_i
+ |D_\mu \tilde{l}^{\prime}_i|^2 +  |D_\mu \tilde{l^c}^{\prime}_i|^2 ~~.
\]
The mass matrices for leptons and sleptons in this basis read:
\begin{equation}
l^c m_l \, l = l^{c \, \prime} R^c (W^c)^{T} m_l W R \, l^{\prime}~~ \equiv
l^{c \, \prime} m_l^\prime l^{\prime},
\end{equation}
and
\footnote{Note that the matrix
$m_{LL}^2$ stands for both, the mass matrix of the left-handed
charged sleptons and the one of the sneutrinos.}
\begin{eqnarray}\nonumber
&& \tilde{l^c} m_{RL}^{2} \, \tilde{l} = \tilde{l^{c}}^{\prime} R^c (W^c)^{T}
m_{RL}^2 W R \, \tilde{l}^{\prime}~~,\\
&& \bar{\tilde{l}} \, m_{LL}^{2} \, \tilde{l} =
\bar{\tilde{l}}^{\prime} \, R W^{\dagger}  \, m_{LL}^2 \, W R \, \tilde{l}^{\prime}~~,\\ \nonumber
&&  \tilde{l^c} \, m_{RR}^{2} \, \bar{\tilde{l^c}} =
\tilde{l^{c \, \prime}} \,  R^c (W^c)^{T} m_{RR}^2 \, W^{c \, \star} R^c \, \bar{\tilde{l^{c \, \prime}}}~~.
\end{eqnarray}
We diagonalize the resulting mass matrix $m_l^\prime$ of the charged leptons
by the usual bi-unitary transformation:
\begin{equation}
U^{T} m_l^\prime V = {\tt diag} (m_e,m_\mu,m_\tau)
\end{equation}
and arrive at the mass eigenbasis $l^{\prime\prime}$ and $l^{c \, \prime\prime}$:
\begin{equation}
l^{c \, \prime} = U l^{c \, \prime\prime} ~~~~~~~~~\mbox{and} ~~~~~~~~~ l^{\prime} = V l^{\prime\prime}~~~.
\end{equation}
Finally, the slepton mass matrices $m_{RL}^{2}$, $m_{LL}^2$ and $m_{RR}^2$
are given as:
\begin{eqnarray}\nonumber
&& \tilde{l}^c m_{RL}^{2} \, \tilde{l} = \tilde{l^c} ^{\prime\prime} [U^{T} R^c (W^c)^{T} m_{RL}^{2} W R
V] \, \tilde{l}^{\prime\prime} \equiv \tilde{l^c} ^{\prime\prime} \hat{m}_{RL}^2 \,
\tilde{l}^{\prime\prime}~~,\\
&& \bar{\tilde{l}} \, m_{LL}^{2} \, \tilde{l} =
\bar{\tilde{l}}^{\prime\prime} [V^{\dagger} R W^{\dagger} \, m_{LL}^2 \, W R V]
\, \tilde{l}^{\prime\prime} \equiv \bar{\tilde{l}}^{\prime\prime} \hat{m}_{LL}^2 \,
\tilde{l}^{\prime\prime}~~,\\ \nonumber
&& \tilde{l^c} \, m_{RR}^{2} \, \bar{\tilde{l^c}} =
\tilde{l^{c \, \prime\prime}} [U^{T} R^c (W^c)^{T} m_{RR}^2 \, W^{c \, \star} R^{c} \, U^{\star}] \,
\bar{\tilde{l^{c \, \prime\prime}}} \equiv \tilde{l^{c \, \prime\prime}} \hat{m}_{RR}^2 \,
\bar{\tilde{l^{c \, \prime\prime}}}~~.
\end{eqnarray}

As we assume for the actual calculation of $\hat{m}_{RL}^2$, $\hat{m}_{LL}^2$ and $\hat{m}_{RR}^2$
 that all couplings involved are
real, the matrices $W$, $W^c$, $U$ and $V$ turn out to be orthogonal instead of unitary. Furthermore,
we express the small parameter $u$ in terms of $t$ as $u = x t$ with $x \leq 1$ according to eq.
(\ref{tbound}) and eq. (\ref{ubound}). We then can do the calculation in just one expansion parameter $t$.
\footnote{The different factors of $t$
and $u$ are recovered in the final result by replacing $x$ with $\frac{u}{t}$.}
In the course of the calculation we pose the following
requirements: the kinetic terms are canonically normalized up to and including $\mathcal{O}(t^5)$,
$m_l^\prime$ is diagonal also up to and including $\mathcal{O}(t^5)$, and the matrices $W$, $W^c$, $U$ and $V$
are orthogonal up to the same order. These calculations have been checked by using two independent methods.


\newcommand{\ijg}{l_i\rightarrow l_j\gamma}
\newcommand{\ijnn}{l_i\rightarrow l_j\nu_i\overline{\nu}_j}
\newcommand{\mtl}{m_{\tilde{l}}}
\newcommand{\mtlx}{m_{\tilde{l}_X}}
\newcommand{\mtn}{m_{\tilde{\nu}}}
\newcommand{\mtnx}{m_{\tilde{\nu}_X}}
\newcommand{\mtca}{M_{\tilde{\chi}_A}}
\newcommand{\mtcn}{M_{\tilde{\chi}^0}}
\newcommand{\mtcna}{M_{\tilde{\chi}^0_A}}
\newcommand{\mtcca}{M_{\tilde{\chi}^-_A}}

\mathversion{bold}
\section*{Appendix C~~~Notations and one-loop formulae for $R_{ij}$ and $\delta a_\mu^{SUSY}$}
\mathversion{normal}

In this part we fix the notations and we report the formulae for $R_{ij}$ and for $\delta a_\mu^{SUSY}$ which we have used in section 5. The main references are \cite{deltaamu_susy,Arganda,HisanoFukuyama,g2BR} and we follow these.

\noindent The mass matrix of the charginos is given by:

\beq
-\cL_m\supset\left(\overline{\widetilde{W}^-_R}\quad\overline{\widetilde{H}^-_{2R}}\right)M_c\left(
         \begin{array}{c}
           \widetilde{W}^-_L \\
           \widetilde{H}^-_{1L} \\
         \end{array}
       \right)+h.c.
\eeq
with
\beq
M_c=
\left(
  \begin{array}{cc}
    M_2 & \sqrt{2}m_W\cos{\beta} \\
    \sqrt{2}m_W\sin{\beta} & \mu \\
  \end{array}
\right)\;.
\eeq
This matrix is diagonalized by $2\times 2$ rotation matrices $O_L$ and $O_R$ as:
\beq
O_R M_c O_L^T={\tt diag} \left(M_{ \widetilde{\chi}^-_1}, M_{ \widetilde{\chi}^-_2}\right)\;,
\eeq
where the diagonalizing matrices connect mass and interaction eigenstates in the following way:
\beq
\left(
 \begin{array}{c}
   \widetilde{\chi}^-_{1L} \\
   \widetilde{\chi}^-_{2L} \\
 \end{array}
\right)=O_L\left(
             \begin{array}{c}
               \widetilde{W}^-_L \\
               \widetilde{H}^-_{1L} \\
             \end{array}
            \right) ~~, ~~~
\left(
 \begin{array}{c}
   \widetilde{\chi}^-_{1R} \\
   \widetilde{\chi}^-_{2R} \\
 \end{array}
\right)=O_R\left(
             \begin{array}{c}
               \widetilde{W}^-_R \\
               \widetilde{H}^-_{2R} \\
             \end{array}
            \right)
\eeq
and the mass eigenstates are written as $\widetilde{\chi}^-_A= \widetilde{\chi}^-_{AL}+ \widetilde{\chi}^-_{AR}$ ($A=1,2$) with masses $M_{ \widetilde{\chi}^-_A}$.

\noindent The neutralino mass matrix is given by:
\beq
-\cL_m\supset\dfrac{1}{2}\left(\begin{array}{cccc}\widetilde{B}_L&\widetilde{W}^0_L&
\widetilde{H}^0_{1L}&\widetilde{H}^0_{2L}\end{array}\right)M_N
\left(
 \begin{array}{c}
   \widetilde{B}_L \\
   \widetilde{W}^0_L \\
   \widetilde{H}^0_{1L} \\
   \widetilde{H}^0_{2L} \\
 \end{array}
\right)+h.c.
\eeq
with
\beq
M_N=\left(
  \begin{array}{cccc}
    M_1 & 0 & -m_Z\sin{\theta_W}\cos{\beta} & m_Z\sin{\theta_W}\sin{\beta} \\
    0 & M_2 & m_Z\cos{\theta_W}\cos{\beta} & -m_Z\cos{\theta_W}\sin{\beta} \\
    -m_Z\sin{\theta_W}\cos{\beta} & m_Z\cos{\theta_W}\cos{\beta} & 0 & -\mu \\
    m_Z\sin{\theta_W}\sin{\beta} & -m_Z\cos{\theta_W}\sin{\beta} & -\mu & 0 \\
  \end{array}
\right)
\label{MN}
\eeq
We can diagonalize $M_N$ by a rotation matrix $O_N$:
\beq
O_N M_N O_N^T={\tt diag} \left(M_{ \widetilde{\chi}^0_1}, M_{ \widetilde{\chi}^0_2}, M_{ \widetilde{\chi}^0_3}, M_{ \widetilde{\chi}^0_4}\right)\;,
\eeq
where $O_N$ connects mass and interaction eigenstates in the following way:
\beq
\left(
 \begin{array}{c}
   \widetilde{\chi}^0_{1L} \\
   \widetilde{\chi}^0_{2L} \\
   \widetilde{\chi}^0_{3L} \\
   \widetilde{\chi}^0_{4L} \\
 \end{array}
\right)=O_N\left(
 \begin{array}{c}
   \widetilde{B}_L \\
   \widetilde{W}^0_L \\
   \widetilde{H}^0_{1L} \\
   \widetilde{H}^0_{2L} \\
 \end{array}
\right)
\eeq
and the mass eigenstates are given by $\widetilde{\chi}^0_A= \widetilde{\chi}^0_{AL}+ \widetilde{\chi}^0_{AR}$ ($A=1,2,3,4$) with masses $M_{ \widetilde{\chi}^0_A}$.\\
The mass matrices for the charged sleptons and for sneutrinos are given by:
\beq
-\cL_m\supset\left(\bar{\tilde{l}}\quad\tilde{l}^c\right)
\hat{\cM}_e^2
\left(
  \begin{array}{c}
    \tilde{l} \\
    \bar{\tilde{l}}^{c} \\
  \end{array}
\right)+\bar{\tilde{\nu}}\hat{m}^2_{\nu LL}\tilde{\nu}
\eeq
with
\beq
\hat{\cM}_e^2=\left(
  \begin{array}{cc}
    \hat{m}^2_{eLL} & \hat{m}_{eLR}^2 \\
    \hat{m}_{eRL}^2 & \hat{m}^2_{e RR} \\
  \end{array}
\right)
\eeq
where $\hat{m}^2_{(e,\nu)LL}$ and $\hat{m}^2_{eRR}$ are in general hermitian matrices and $\hat{m}^2_{eLR}=\left(\hat{m}^2_{RL}\right)^\dag$.
We diagonalize the mass matrix $\hat{\cM}_e^2$ by a $6\times 6$ rotation matrix $U^{\tilde{l}}$ as:
\beq
U^{\tilde{l}}\hat{\cM}_e^2U^{\tilde{l}\;T}=\mtl^2
\eeq
where the mass eigenstates are:
\beq
\tilde{l}_X=U^{\tilde{l}}_{X,i}\tilde{l}_i+U^{\tilde{l}}_{X,i+3}\bar{\tilde{l}}^{c}_i
\eeq
with masses $m^2_{\tilde{l}_X}$ ($X=1,\ldots,6$).

\noindent Analogously, the sneutrino mass matrix is diagonalized by:
\beq
U^{\tilde{\nu}} \hat{m}_{\nu LL}^2U^{\tilde{\nu} T}=\mtn^2
\eeq
where the mass eigenstates are:
\beq
\tilde{\nu}_X=U^{\tilde{\nu}}_{X,i}\tilde{\nu}_i
\eeq
with masses $m^2_{\tilde{\nu}_X}$ ($X=1,2,3$).\\
\\
The normalized branching ratios, $R_{ij}$, for the LFV transitions $\ijg$ are:
\beq
R_{ij}=\dfrac{BR(\ijg)}{BR(\ijnn)}=\dfrac{48\pi^3\al}{G_F^2}\left(|A_2^L|^2+|A_2^R|^2\right)
\eeq
and the decay rates are given by:
\beq
\Ga(\ijnn)=\dfrac{G_F^2}{192\pi^3}m_i^5 ~~, ~~~
\Gamma(\ijg)=\dfrac{e^2}{16\pi}m_i^5\left(|A_2^L|^2+|A_2^R|^2\right)\;.
\eeq
Each coefficient $A_2^{L,R}$ can be written as a sum of two terms:
\be
A_2^{L,R}=A_2^{(n)L,R}+A_2^{(c)L,R}\;,
\ee
where $A_2^{(n)L,R}$ and $A_2^{(c)L,R}$ stand for the contributions from the neutralino and from the chargino loops, respectively.
These coefficients are explicitly given by:
\beq
A_2^{(n)L}=\dfrac{1}{32\pi^2}\dfrac{1}{\mtlx^2}\bigg[N_{jAX}^L \bar{N}_{iAX}^{L} g_{1n}(x_{AX})+N_{jAX}^R \bar{N}_{iAX}^{R}\dfrac{m_j}{m_i} g_{1n}(x_{AX})
  +N_{jAX}^L \bar{N}_{iAX}^{R}\dfrac{\mtcna}{m_i} g_{2n}(x_{AX})\bigg]
\eeq
and $A_2^{(n)R}=A_2^{(n)L}\vert_{L\leftrightarrow R}$ with $x_{AX}=\mtcna^2 / \mtlx^2$, and 
\beq
A_2^{(c)L}=-\dfrac{1}{32\pi^2}\dfrac{1}{\mtnx^2}\bigg[C_{jAX}^L \bar{C}_{iAX}^{L} g_{1c}(x_{AX})+C_{jAX}^R \bar{C}_{iAX}^{R}\dfrac{m_j}{m_i} g_{1c}(x_{AX})
 +C_{jAX}^L \bar{C}_{iAX}^{R}\dfrac{\mtcca}{m_i} g_{2c}(x_{AX})\bigg]
\eeq
and $A_2^{(c)R}=A_2^{(c)L}\vert_{L\leftrightarrow R}$ with $x_{AX}=\mtcca^2 / \mtnx^2$.\\
The terms $N_{iAX}$ and $C_{iAX}$ and the loop functions $g_{in}$ and $g_{ic}$ read as follows:
\begin{eqnarray}
&&N_{iAX}^L=-\dfrac{g_2}{\sqrt{2}}\left\{\left[-(O_N)_{A,2}-(O_N)_{A,1}\tan{\theta_W}\right]
U^{\tilde{l}}_{X,i}+\dfrac{m_i}{m_W\cos{\beta}}(O_N)_{A,3}U^{\tilde{l}}_{X,i+3}\right\}\label{functions:NiAXR}\\
[-0.3cm]\nonumber\\
&&N_{iAX}^R=-\dfrac{g_2}{\sqrt{2}}\left\{\dfrac{m_i}{m_W\cos{\beta}}(O_N)_{A,3}U^{\tilde{l}}_{X,i}+ 2(O_N)_{A,1}\tan{\theta_W}U^{\tilde{l}}_{X,i+3}\right\}\label{functions:NiAXL}\\
[-0.3cm]\nonumber\\
&&C_{iAX}^L=-g_2(O_R)_{A,1}U^{\tilde{\nu}}_{X,i}\label{functions:CiAXR}\\
[-0.3cm]\nonumber\\
&&C_{iAX}^R=g_2\dfrac{m_i}{\sqrt{2}m_W\cos{\beta}}(O_L)_{A,2}U^{\tilde{\nu}}_{X,i}\label{functions:CiAXL}
\end{eqnarray}
and
\begin{eqnarray}
&&g_{1n}(x_{AX})=\dfrac{1}{6(1-x_{AX})^4}\left(1-6x_{AX}+3x_{AX}^2+2x_{AX}^3-6x_{AX}^2\ln{x_{AX}}\right)\label{functions:g1n}\\
[-0.3cm]\nonumber\\
&&g_{2n}(x_{AX})=\dfrac{1}{(1-x_{AX})^3}\left(1-x_{AX}^2+2x_{AX}\ln{x_{AX}}\right)\label{functions:g2n}\\
[-0.3cm]\nonumber\\
&&g_{1c}(x_{AX})=\dfrac{1}{6(1-x_{AX})^4}\left(2+3x_{AX}-6x_{AX}^2+x_{AX}^3+6x_{AX}\ln{x_{AX}}\right)\label{functions:g1c}\\
[-0.3cm]\nonumber\\
&&g_{2c}(x_{AX})=\dfrac{1}{(1-x_{AX})^3}\left(-3+4x_{AX}-x_{AX}^2-2\ln{x_{AX}}\right)\label{functions:g2c}\;.
\end{eqnarray}\\
We note that the functions $f_{i n}$ and $f_{i c}$ , displayed in section 4, are related to the loop functions $g_{in}$ and  $g_{ic}$, mostly
through taking the first derivative.

The deviation of the muon anomalous magnetic moment, $\delta a_\mu^{SUSY}$, due to SUSY contributions can be written as:
\beq
\delta a_{\ell_i}^{SUSY}=\dfrac{g_{\ell_i}^{(n)}+g_{\ell_i}^{(c)}}{2}
\eeq
with
\beq
g_{\ell_i}^{(n,c)}=g_{\ell_i}^{(n,c)L}+g_{\ell_i}^{(n,c)R}\;.
\eeq
These terms are explicitly given by:
\beq
g_{\ell_i}^{(c)L}=\dfrac{1}{16\pi^2}\dfrac{m_i^2}{\mtnx^2}\bigg[C_{iAX}^L \bar{C}_{iAX}^{L} g_{1c}(x_{AX})+C_{iAX}^R \bar{C}_{iAX}^{R} g_{1c}(x_{AX})
 +C_{iAX}^L \bar{C}_{iAX}^{R}\dfrac{\mtcca}{m_i} g_{2c}(x_{AX})\bigg]
\eeq
and $g_{\ell_i}^{(c)R}=g_{\ell_i}^{(c)L}\vert_{L\leftrightarrow R}$ with $x_{AX}=\mtcca^2/\mtnx^2$, and 
\beq
g_{\ell_i}^{(n)L}=-\dfrac{1}{16\pi^2}\dfrac{m_i^2}{\mtlx^2}\bigg[N_{iAX}^L \bar{N}_{iAX}^{L} g_{1n}(x_{AX})+N_{iAX}^R \bar{N}_{iAX}^{R} g_{1n}(x_{AX})
+N_{iAX}^L \bar{N}_{iAX}^{R}\dfrac{\mtcna}{m_i} g_{2n}(x_{AX})\bigg]
\eeq
and $g_{\ell_i}^{(n)R}=g_{\ell_i}^{(n)L}\vert_{L\leftrightarrow R}$ with $x_{AX}=\mtcna^2 / \mtlx^2$.
The terms $N_{iAX}$ and $C_{iAX}$ and the functions $g_{in}$ and $g_{ic}$
have been already introduced in eqs. (\ref{functions:NiAXR}-\ref{functions:CiAXL}) and in eqs. (\ref{functions:g1n}-\ref{functions:g2c}).


\end{document}